\documentclass[%
 reprint,%
 amssymb, amsmath,%
aps,
]{revtex4-1}

\usepackage{graphics}
\usepackage{graphicx}
\usepackage{amsmath}
\usepackage{amsfonts}
\usepackage{latexsym}
\usepackage{amsbsy}
\usepackage{epstopdf}
\usepackage{bm}%

\newcommand{\ket}[1]{\vert #1 \rangle}

\usepackage{color}

\usepackage{ulem} 

\newcommand{\ef}{E-field}
\newcommand{\efs}{E-fields}
\newcommand{\efm}{E-field measurement}
\newcommand{\efms}{E-field measurements}

\newcommand{\dyadic}[1]{{#1}
\setbox0=\hbox{$\scriptstyle\leftrightarrow$}
   \setbox2=\hbox{$#1$}
   \dimen0=.5\wd0 \advance\dimen0 by-.5\wd2
   \advance\dimen0 by-\wd0
   \kern\dimen0
{^{\hbox{$\scriptstyle\leftrightarrow$}}}}

\begin{document}

\title{Electromagnetically Induced Transparency (EIT) and Autler-Townes (AT) splitting in the Presence of Band-Limited White Gaussian Noise}
\thanks{Publication of the U.S. government, not subject to U.S. copyright. Approved for Public Release; Distribution Unlimited. Case Number 17-4520. \copyright2017 The MITRE Corporation. ALL RIGHTS RESERVED.}
\author{Matthew T. Simons}
\affiliation{National Institute of Standards and Technology (NIST), Boulder,~CO~80305, USA}
\author{Marcus D. Kautz}
\affiliation{National Institute of Standards and Technology (NIST), Boulder,~CO~80305, USA}
\author{Christopher~L.~Holloway}
\email{holloway@boulder.nist.gov}
\affiliation{National Institute of Standards and Technology (NIST), Boulder,~CO~80305, USA}
\author{David A. Anderson}
\affiliation{Rydberg Technologies, LLC, Ann Arbor, MI 48104, USA}
\author{Georg Raithel}
\affiliation{Rydberg Technologies, LLC, Ann Arbor, MI 48104, USA}
\affiliation{Department of Physics, University of Michigan, Ann Arbor, MI 48109, USA}
\author{Daniel Stack}
\affiliation{The MITRE Corporation, Princeton, NJ 08540, USA}
\author{Marc C. St. John}
\affiliation{The MITRE Corporation, McLean, VA 22102, USA}
\author{Wansheng Su}
\affiliation{The MITRE Corporation, McLean, VA 22102, USA}

\date{\today}

\begin{abstract}
We investigate the effect of band-limited white Gaussian noise (BLWGN) on electromagnetically induced transparency (EIT)
and Autler-Townes (AT) splitting, when performing atom-based continuous-wave (CW) radio-frequency (RF) electric (E) field strength measurements with Rydberg atoms in an atomic vapor. This EIT/AT-based E-field measurement approach is currently being investigated by several groups around the world as a means to develop a new SI traceable RF E-field measurement technique. For this to be a useful technique, it is important to understand the influence of BLWGN. We perform EIT/AT based E-field experiments with BLWGN centered on the RF transition frequency and for the BLWGN blue-shifted and red-shifted relative to the RF transition frequency. The EIT signal can be severely distorted for certain noise conditions (band-width, center-frequency, and noise power), hence altering the ability to accurately measure a CW RF E-field strength. We present a model to predict the changes in the EIT signal in the presence of noise. This model includes AC Stark shifts and on resonance transitions associated with the noise source. The results of this model are compared to the experimental data and we find very good agreement between the two.
\end{abstract}

\maketitle

\section{Introduction}

Significant progress has been made in the development of a novel Rydberg-atom spectroscopic approach for radio-frequency (RF) electric (E) field strength measurements \cite{r1, r2, r3, r5, r6a, r7, r8, r8b, fan2, dave1, dave2}.  This approach utilizes the phenomena of electromagnetically induced transparency (EIT) and Autler-Townes (AT) splitting \cite{r1,r2, r6a, EIT_Adams}, and can lead to a direct International System of Units (SI) traceable, self-calibrated measurement. For the
method to be accepted by National Metrology Institutes as a new international standard for E-field measurements and calibrations, various aspects of the measurement approach must be investigated. One key issue is the ability of this EIT/AT-base technique to measure an RF E-field in the presence of noise. Here, we perform experiments measuring RF E-field strengths in the presence of band-limited white Gaussian noise (BLWGN).

\begin{figure}[!t]
\centering
\scalebox{.31}{\includegraphics*{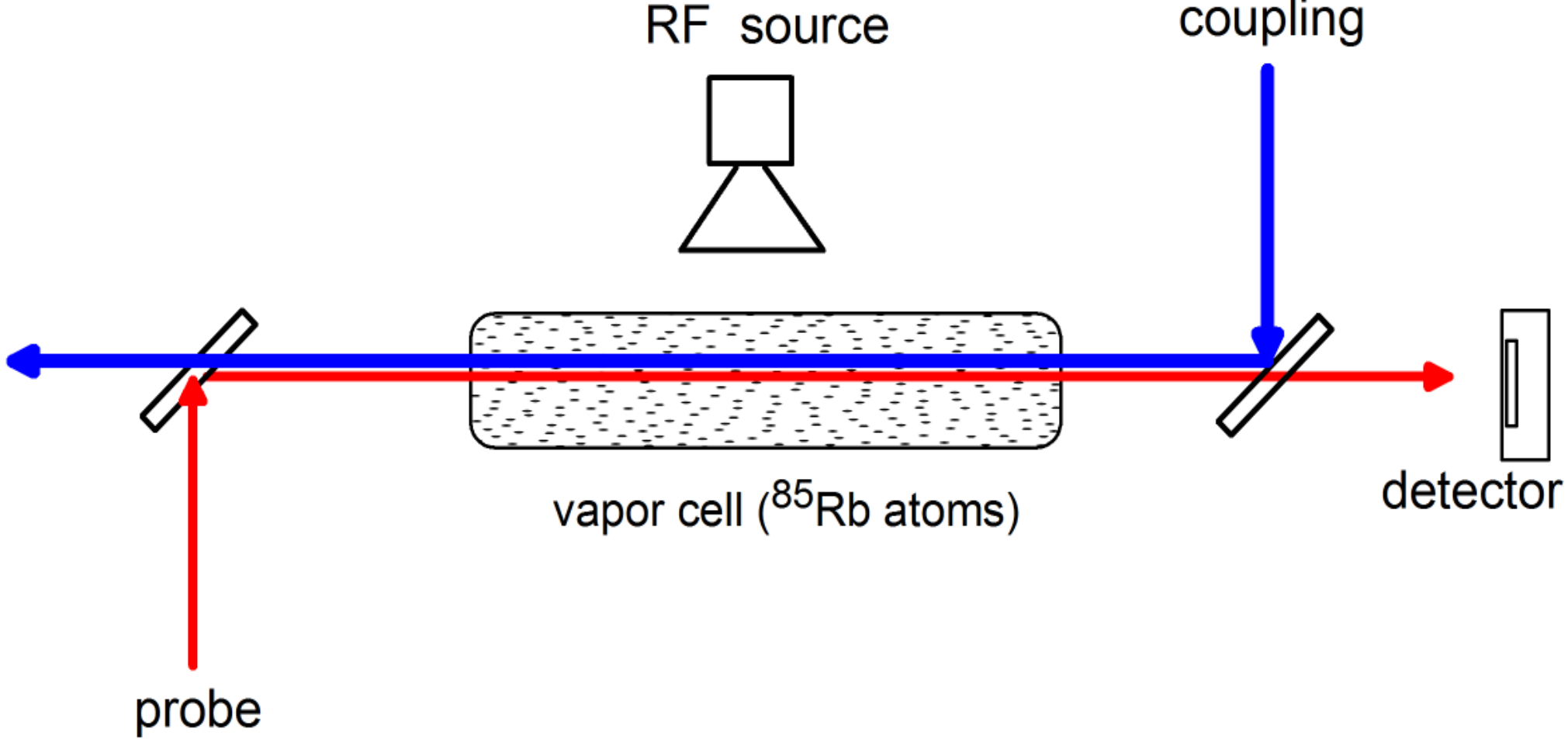}}
\caption{Illustration of the vapor cell setup for measuring EIT, with counter-propagating probe and coupling beams. The RF is applied transverse to the optical beam propagation in the vapor cell.}
\label{4level}
\end{figure}

\begin{figure}[!t]
\centering
\scalebox{.15}{\includegraphics*{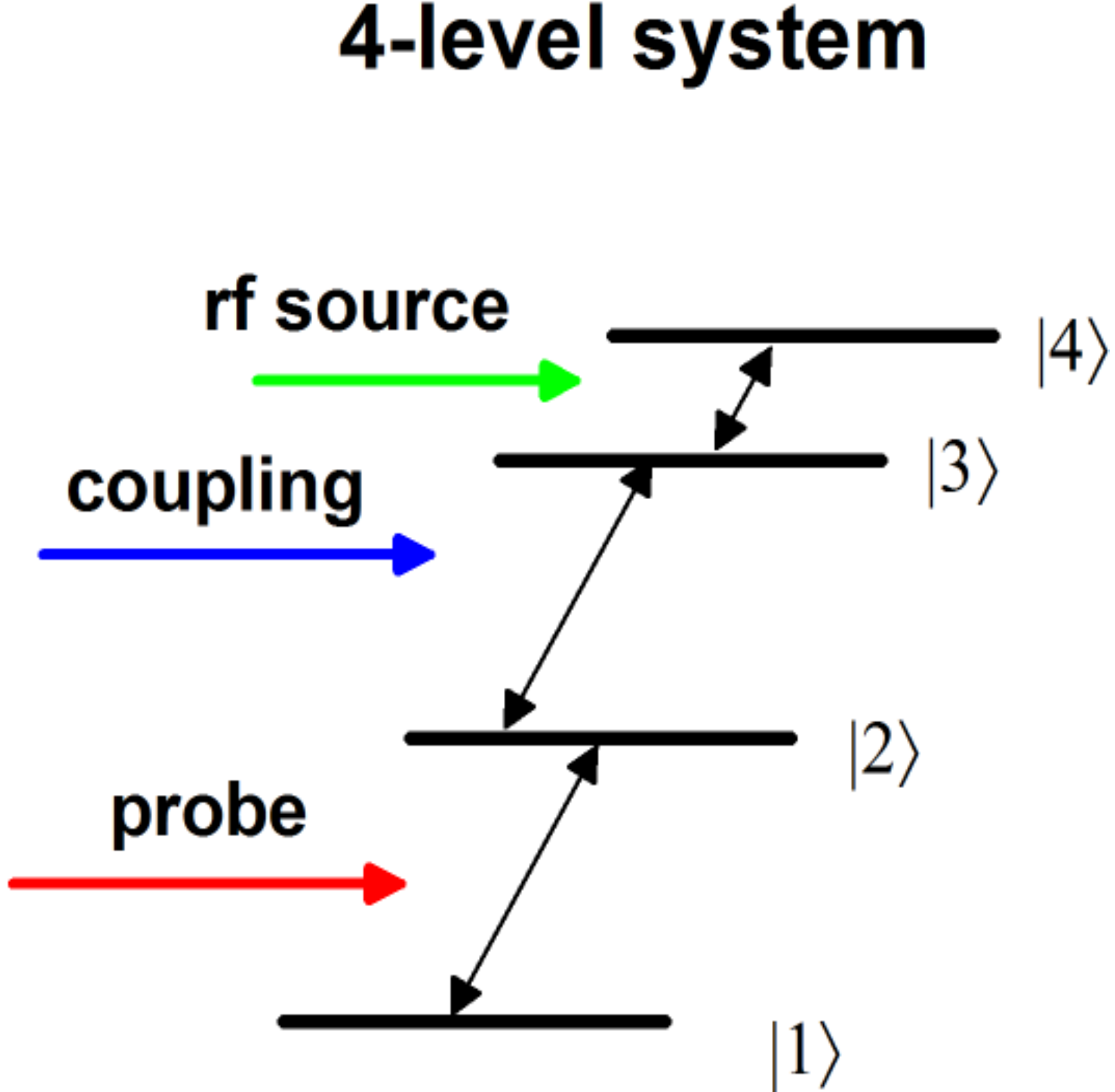}}\hspace{10mm}
\scalebox{.2}{\includegraphics*{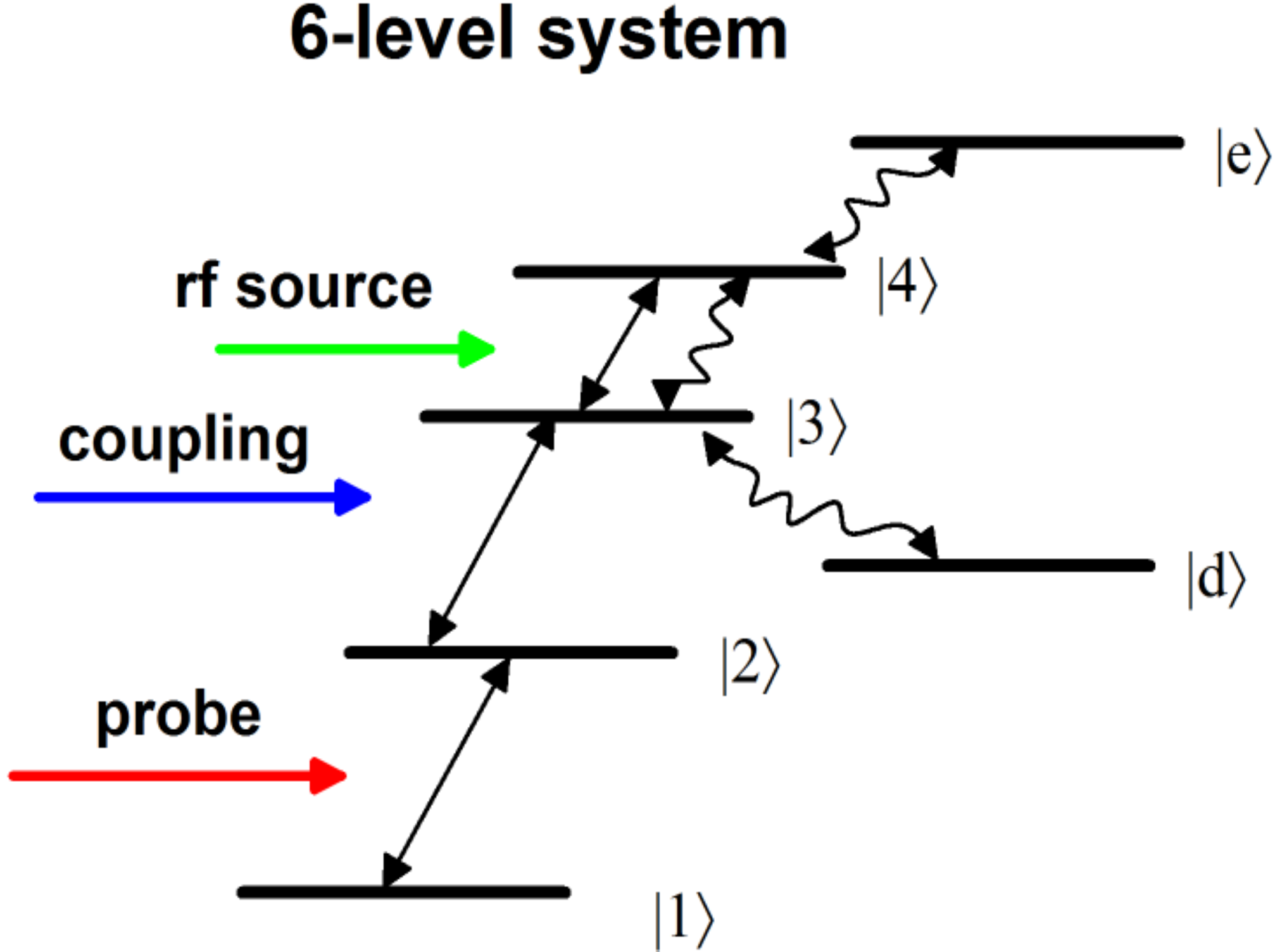}}\\
{\hspace{-5mm}\tiny{(a) \hspace{35mm} (b)}}\\
\caption{Illustration of the atomic systems describing the measurement: (a) a four-level system when no noise is present and (b) a six-level system when noise is present. Coherent transitions are indicated by ``$\leftrightarrow$'' and noise induced transitions and shifts are indicated by ``$\leftrightsquigarrow$''.}
\label{4level2}
\end{figure}

\begin{figure}[!t]
\centering
\scalebox{.31}{\includegraphics{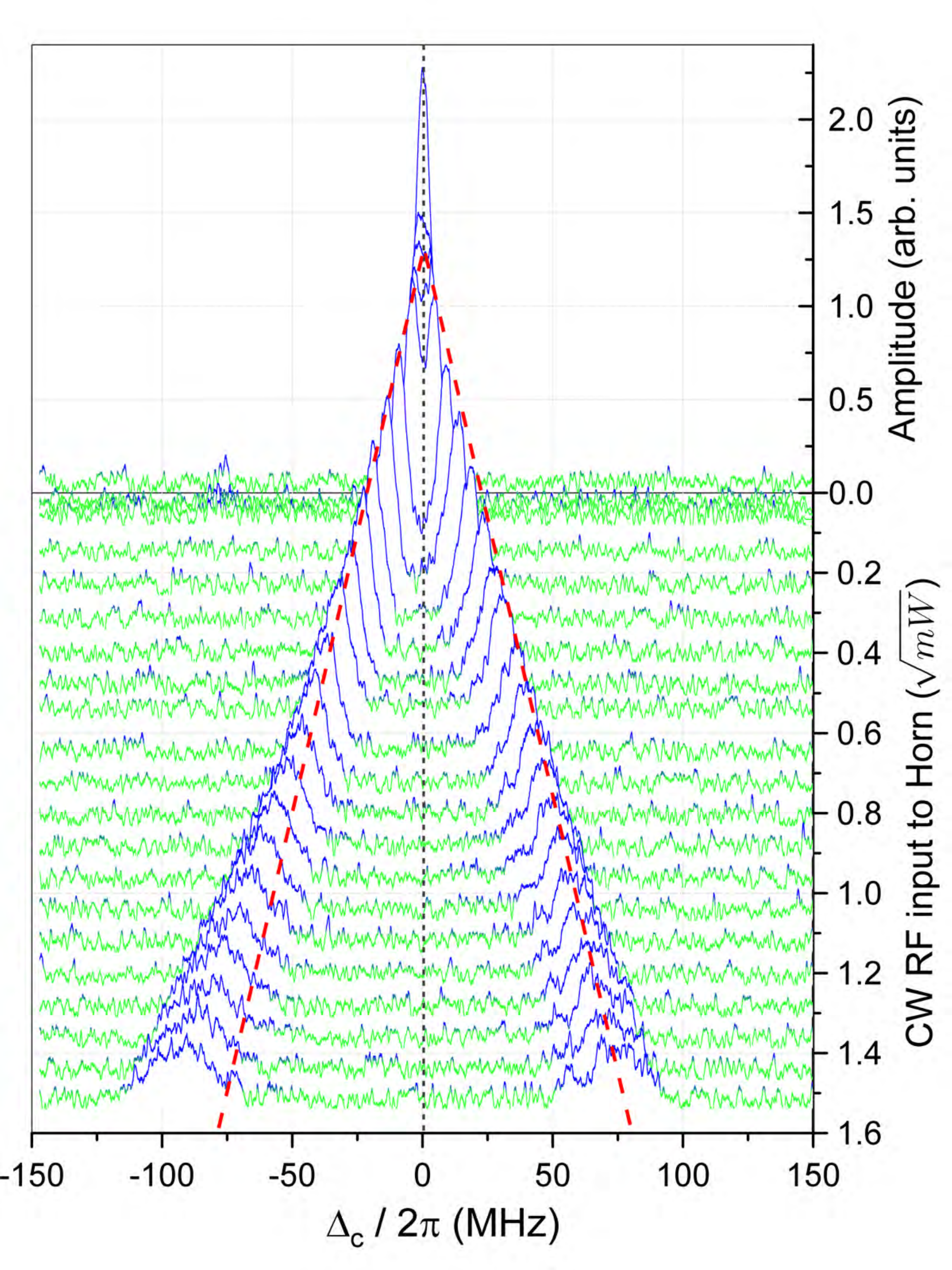}}
\caption{Experimental data for the EIT signal obtained without noise. The figure shows the probe laser transmission
through the cell as a function of coupling laser detuning $\Delta_c$ and for different RF E-field strengths. This dataset is for
a RF of 19.7825~GHz and corresponds to the following $^{85}$Rb 4-level atomic system: 5$S_{1/2}$-5$P_{3/2}$-57$S_{1/2}$-57$P_{1/2}$.}
\label{EIT}
\end{figure}

The measurement approach used to measure the RF E-field strength when no noise is present can be explained by the schematic and the four-level atomic system shown in Figs. \ref{4level} and \ref{4level2}(a). (Note that when noise is present, the six-level atomic system shown in Fig.~\ref{4level2}(b) is required, which is explained below.) A ``probe'' laser is used to probe the response of the ground-state transition of the atoms, and a second laser (``coupling'' laser) is used to excite the atoms to a Rydberg state. In the presence of the coupling laser, a destructive
quantum interference occurs and the atoms become transparent to the resonant probe laser. This is the concept of EIT, in which a transparency window is opened for the probe laser light: probe light transmission is increased.  The coupling laser wavelength
is chosen such that the atom is in a sufficiently high state (a Rydberg state) such that a radio frequency (RF) field coherently couples two Rydberg states
(levels 3 and 4 in Fig. \ref{4level2}(a)).  The RF field in the four-level atomic system causes constructive interference of excitation
pathways within the EIT transmission window, resulting in a decreased transmission of the probe laser and AT splitting
of the EIT peak. A typical measured spectrum for an RF source with different power levels is shown in Fig.~\ref{EIT}. This figure shows the measured EIT signal for a range of E-field strengths (more details on these results are given below). In this figure, $\Delta_c$ is the detuning of the coupling laser  (where $\Delta_c=\omega_c-\omega_o$; $\omega_o$ is the
on-resonance angular frequency of the Rydberg state transition and $\omega_c$ is the angular frequency of the coupling laser). Notice that the AT splitting increases with increasing applied E-field strength. Here, we explore how such measurements are affected by noise-induced transitions and level shifts introduced by BLWGN.

Under the absence of BLWGN, the AT splitting (defined as $2\pi\Delta f_o$) of the coupling laser spectrum is easily measured and under certain conditions is equal
to the Rabi frequency of the RF transition \cite{linear},
\begin{equation}
{\rm AT\,\,\,\, splitting}=\Omega_{RF}=2\pi \Delta f_o \,\,\, ,
\label{foeq}
\end{equation}
where $\Omega_{RF}=|E|{\wp}/{\hbar}$ is the Rabi frequency of the RF transition, $\hbar$ is Planck's constant,
and $\wp$  is the dipole moment of the atomic RF transition. This relationship between the AT splitting and
the Rabi frequency is obtained in the weak probe limit and for no Doppler averaging.  By measuring this splitting
($\Delta f_m$) we get a direct measurement of the RF E-field strength. In this approach, either the probe or the coupling laser can be scanned or detuned. For either case, the E-field strength is given by~\cite{r1, r2},
\begin{equation}
|E| = 2 \pi \frac{\hbar}{\wp} D\, \Delta f_m= 2 \pi \frac{\hbar}{\wp}\Delta f_0 \,\,\, ,
\label{mage}
\end{equation}
where $\Delta f_m$ is the measured splitting, $\Delta f_o=D\, \Delta f_m$, and $D$ is a parameter whose value depends on which of the two lasers is scanned during the measurement. If the probe laser is scanned, $D=\frac{\lambda_p}{\lambda_c}$, where $\lambda_p$ and $\lambda_c$ are the wavelengths of the probe and coupling laser, respectively. This ratio is needed to account for the Doppler mismatch of the probe and coupling lasers \cite{EIT_Adams}. If the coupling laser is scanned, it is not required to correct for the Doppler mismatch, and $D=1$. This type of measurement of the E-field strength is considered a direct SI-traceable, self-calibrated measurement because it is directly related to Planck's constant (which will become an SI-defined quantity by standards bodies in the near future), the atomic dipole moment $\wp$ (a parameter which can be calculated very accurately \cite{r1, r6}), and only requires a relative optical frequency measurement $\Delta f_m$, which can be measured very accurately.

To investigate how the EIT signals shown in Fig.~\ref{EIT} are influenced by the presence of BLWGN, and in turn the ability to measure an E-field strength, we perform experiments with four BLWGN sources: (1) BLWGN with a center frequency above the frequency of the coherently driven transition (blue-shifted BLWGN), (2) BLWGN centered about the RF transition frequency, (3) BLWGN with a center frequency below the RF transition frequency (red-shifted BLWGN), and (4) BLWGN with a notch around the RF transition (i.e., there is a noise band above and below the RF transition resonant frequency, but no noise at the RF transition frequency).  We present a model to predict the observed behavior of the EIT signal in the presence of BLWGN, and show excellent agreement between experimental and model results.


\section{Experimental Setup and Noise Source}
\begin{figure}
\centering
\scalebox{.1}{\includegraphics{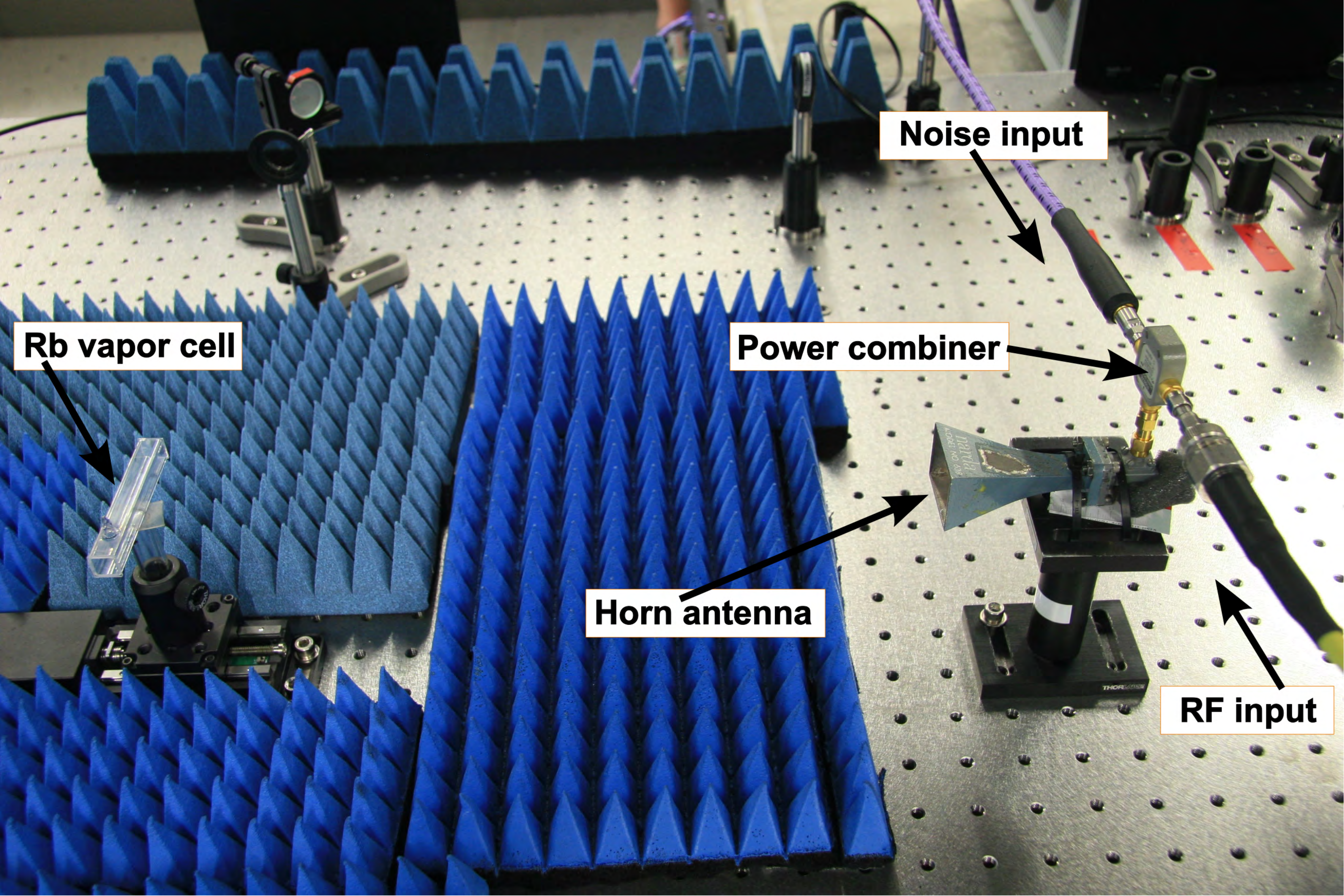}}
\scriptsize{\centerline{(a) photo of experimental setup}}\\
\vspace{4mm}
\scalebox{.4}{\includegraphics*{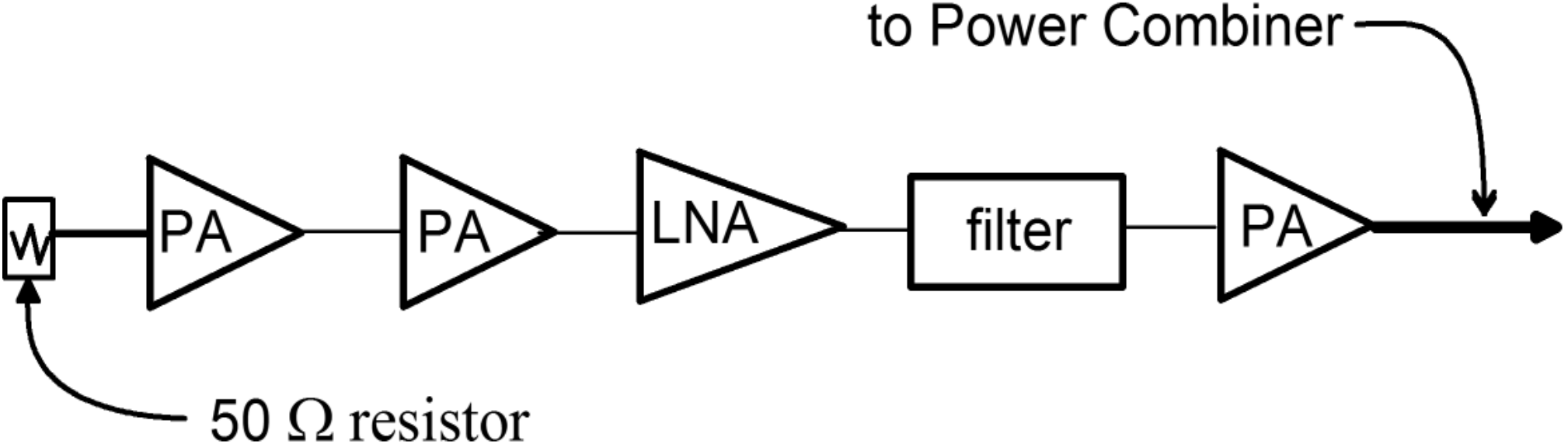}}
\scriptsize{\centerline{(b) block diagram}}
\caption{Experimental setup for E-field measurements using EIT: a) picture of the setup and (b) block diagram of the noise source setup. RF absorber (blue pyramidal cones in the photo) is used to reduce RF reflections from the table and optical components. }
\label{layout}
\end{figure}

A picture of the experimental setup and a block diagram of the noise source are shown in Fig.~\ref{layout}. The atom-based measurements are done using a $10~\times~10~\times~75~$mm rectangular vapor cell filled with $^{85}$Rb, two lasers (a probe and a coupling laser), a photo-detector, and a lock-in amplifier. A diagram of the laser orientations in the vapor cell is shown in Fig.~\ref{4level}. The levels $\ket{1}$, $\ket{2}$, $\ket{3}$, and $\ket{4}$ in Fig.~\ref{4level2}(a) correspond respectively to the $^{85}$Rb  $5S_{1/2}$ ground state,  $5P_{3/2}$ excited state, and two Rydberg states $57{\rm S}_{1/2}$ and $57{\rm P}_{1/2}$. The probe is a $780.24~$nm laser focused inside the cell to a full-width at half maximum (FWHM) of $270~\mu$m, and has a power of $4.1~\mu$W.  To produce an EIT signal, we applied a counter-propagating coupling laser (which is overlapped with the probe laser) tuned to $479.9285$~nm to couple the $5{\rm P}_{3/2}$ and $57{\rm S}_{1/2}$ states, with a FWHM of $353~\mu$m and power of $43.3~$mW. A $19.7825~$GHz E-field is applied via a Narda 638 standard gain horn antenna (mentioning this product does not imply an endorsement, but serves to clarify the antenna used) to couple the Rydberg states $57{\rm S}_{1/2}-57{\rm P}_{1/2}$. We modulated the coupling laser amplitude with a 50/50 duty-cycle $30~$kHz square wave and detected the resulting modulated probe transmission with a lock-in amplifier to obtain an amplified EIT signal.

A power combiner is connected to the input of the horn antenna to combine the noise signal and the continuous-wave (CW) $19.7825~$GHz signal from a signal generator (SG), such that both noise and CW signals can be incident on the vapor cell simultaneously. The horn antenna is placed $34.2~$cm from the center of the two overlapped laser beams inside the vapor cell.


The noise signal is generated by connecting a $50~\Omega$ resistor to a series of amplifiers, as shown in Fig.~\ref{layout}(b). The resistor is connected in series to two power amplifier (PA) with a gain of 26~dB, and a low low-noise amplifier (LNA) with a gain of 27~dB.  The output of the LNA is sent to a band-pass filter (which was changed to the different bands during the experiment). The output of the filter was then fed into a third PA with $30~$dB gain. The output of the third amplifier was connected to the power combiner shown in Fig.~\ref{layout}(a).

\begin{figure}[!h]
\centering
\scalebox{.3}{\includegraphics{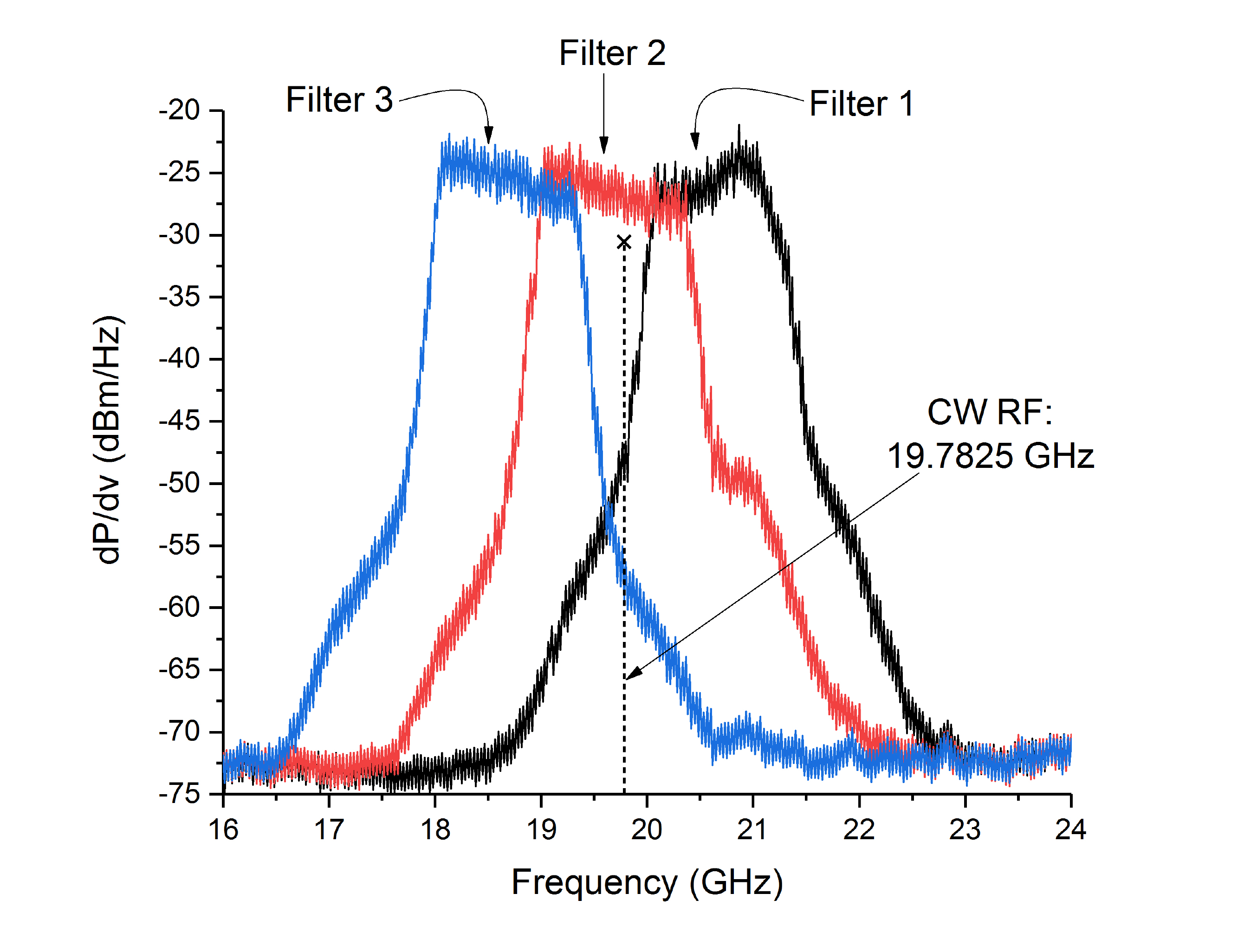}}
\caption{Measured noise power spectral density ($dP/dv$) for the three filters used in the experiments. These data is measured with a spectrum analyzer at the output of the power combiner that feeds the horn antenna. On this plot, we have indicated the CW source frequency.}
\label{noise}
\end{figure}

To exhibit the importance of the detailed noise spectrum on the EIT signal, several different bandpass filters are used to band-limit the noise signal. In these experiments we used three different filters, each with a bandwidth of $\sim1~$GHz, with different center frequencies as follows: Filter 1 $\approx 20.7~$GHz (blue-shifted BLWGN), Filter 2  $\approx 19.7~$GHz (on-resonance BLWGN), and Filter 3 $\approx 18.7~$GHz (red-shifted BLWGN).  Fig.~\ref{noise} shows the noise power spectral density ($dP/dv$) of the band-pass filters, measured with a spectrum analyzer connected to the output of the power combiner (i.e., the input to the horn antenna). Using a power meter, we measured the integrated noise power (total power over the filter bandwidth) for each filter, measured at the output of the power combiner that feeds the horn antenna. The integrated power was $5.4~$dBm for Filter 1, $6.0~$dBm for Filter 2, and $6.6~$dBm for Filter 3.  The average noise power was intentionally set to be approximately the same for all three filters. The fourth type of BLWGN was created by combining Filters 1 and 3 (`Filter 1/3' - noise bands above and below the transition frequency, with a notch on resonance). Using a power splitter, the power output from the LNA was split, with one channel sent through Filter 1 and the other through Filter 3. These two channels were then recombined with a power combiner, and sent to the last PA. This configuration results in a noise spectrum spanning $18.2~$GHz to $21.2~$GHz, with a notch from 19.2-to-20.2~GHz. The integrated noise power for Filter~1/3 was $4.95~$dBm. In this configuration we added additional attenuation to ensure the integrated noise power was approximately the same as for the measurements with a single filter.


\section{Noise Effects on E-field Measurements}

\begin{figure}
	\centering
\scalebox{.28}{\includegraphics{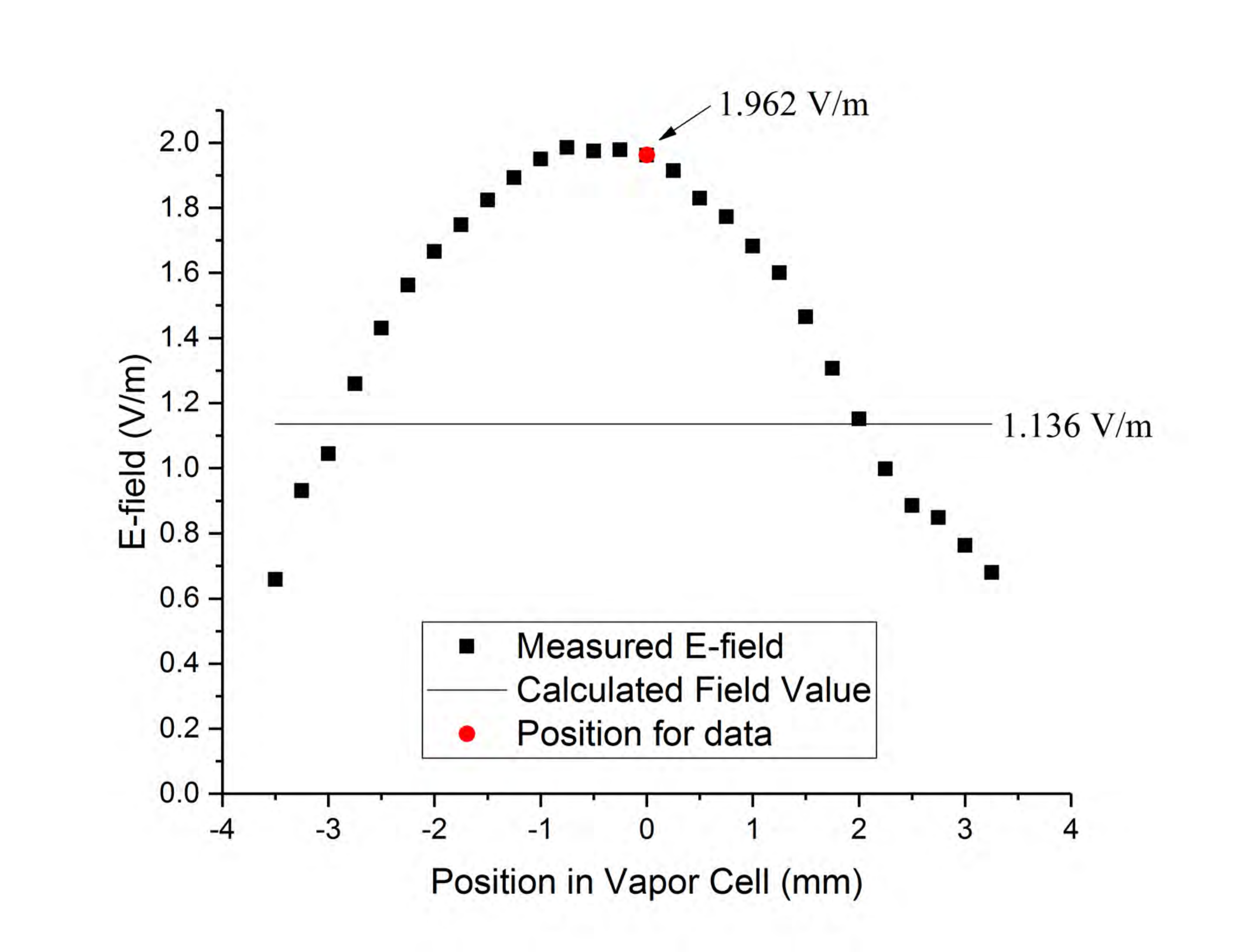}}
	\caption{Measured E-field as a function of position in vapor cell, where $0=~$edge of cell closest to the horn. The line shows a far-field calculation of the E-field at the cell. All data in the subsequential figures was taken at the point indicated by the red diamond.}
	\label{cell_sweep}
\end{figure}

In the experiments we measured the transmitted probe laser power with a photo-detector, as the coupling laser frequency was swept, for a set  of CW RF power levels of the coherent RF source. The CW RF power was varied by the SG, which was fed to the horn antenna through a cable. The measured coherent input power to the horn antenna ranged from 0~mW to 2.4~mW (including loss in the cable).  By taking into account the gain of the antenna ($G\approx15.7$~dB at $19.78~$GHz) and the distance from the antenna to the lasers ($x=0.342~$m), the E-field strengths seen by the atoms are estimated by \cite{stutzman}
\begin{equation}
|E|=\frac{A_{sw}}{x}\sqrt{\frac{c\,\,\mu_0}{2\pi}}\sqrt{P_{SG}\,\,
10^{\frac{G}{10}}}\,\,\, ,
\label{friis}
\end{equation}
where $c$ is the speed of light {\it in vacuo}, $\mu_0$ is the permeability of free-space, and $P_{SG}$ is the power level at the input to the horn antenna in units of Watts. There is an additional parameter $A_{sw}$ which is called the E-field enhancement factor and is a correction factor of the E-field due to a standing wave formed inside the vapor cell \cite{fan}. Following the method in \cite{r3}, we measured the field versus position in the vapor cell (see Fig.~\ref{cell_sweep}) and compared these with the far-field calculation. At the position inside the cell where the measurements were performed we found an E-field enhancement factor of $A_{sw} = 1.73$.

This gave a range of E-field strengths at the location of the laser beam crossing from $0~$V/m to $11.6~$V/m. Fig.~\ref{EIT} shows the measured EIT signal for this range of RF E-field strengths. Throughout the paper, the data in Fig.~{\ref{EIT} will be referred to as CW RF data because they corresponds to an EIT signal with only a CW RF field at $19.7825~$GHz (i.e., no noise present). These data will be compared to the EIT signal in the presence of various noise profiles. We placed a reference line at the RF-free 479.9285~nm Rydberg resonance (i.e., $\Delta_c=0$) and two additional reference lines on either side.  These two additional lines indicate the linear trend that the AT-splitting would follow if the splitting remained linear with the E-field strength.  We see that for larger E-field strengths the EIT signal begins to deviate from linear behavior, with more deviation for the peaks on the $\Delta_c<0$ side of the EIT spectrum.  This deviation is mainly due to an AC Stark shift, caused by the coherent RF signal for high applied E-field strengths \cite{dave1}. When comparing the EIT signals with and without the RF source, we see that the linewidth of the EIT signals in Fig.~\ref{EIT} has some broadening. This broadening is due to the inhomogeneity of the E-field inside the vapor cell \cite{dave2}.

Adding BLWGN to the RF applied through the horn antenna (via the power combiner) distorts the EIT signal shown in Fig.~\ref{EIT}. The amount of distortion is a strong function of both the frequency band of the noise (which bandpass filter is used) and the amount of noise power. Figs.~\ref{noiseEIT1}-\ref{noiseEIT13} show the measured EIT signal for Filters 1, 2, 3 and the combined Filter 1/3, respectively. In these figures, the plots on the left-hand-side of each figure [i.e., (a), (b), and (c)] correspond to the experimental data and the plot on the right-hand-side [i.e., (d), (e), and (f)] correspond to results from a theoretical model (the model is discussed in the next section). The experimental data and theoretical data are shown side-by-side for ease of comparison later. The plots in these figures correspond to the indicated levels of attenuation of the noise sources (i.e., attenuators placed on the noise signal before feeding the horn antenna).  All the plots in these figures have the $\Delta_c=0$ reference line for easier comparisons to Fig.~\ref{EIT}. It is interesting to note the noise sources either red-shift or blue-shift the EIT signal. Filter 1 blue-shifts the EIT signal, whereas Filter 2 and Filter 3 red-shift the EIT signal. All the frequency shifts increase with increasing noise level. The combination Filter 1/3 blue-shifts the EIT signals. The effects of the Filter 1/3 combination are dominated by the contribution from the noise power spectrum in Filter 1.
We also see that for high noise power levels (i.e., 0~dB attenuation), the noise generated by either Filter 1 or Filter 2 dramatically suppresses the EIT signal. This amount of suppression is not observed for the high-noise power case when Filter 3 is used.

\begin{figure*}
\centering
\scalebox{.27}{\includegraphics{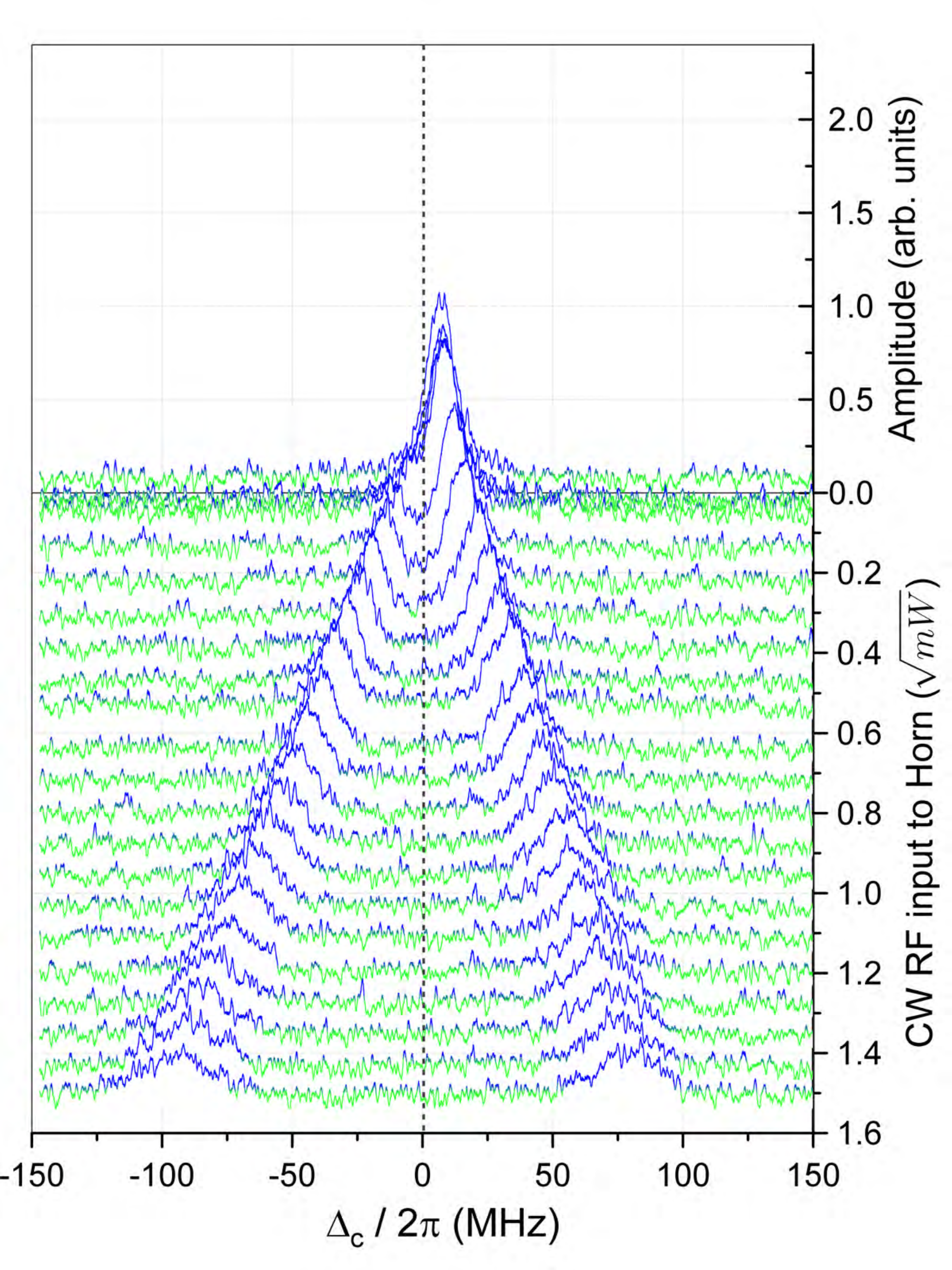}} \hspace{20mm}
\scalebox{.27}{\includegraphics{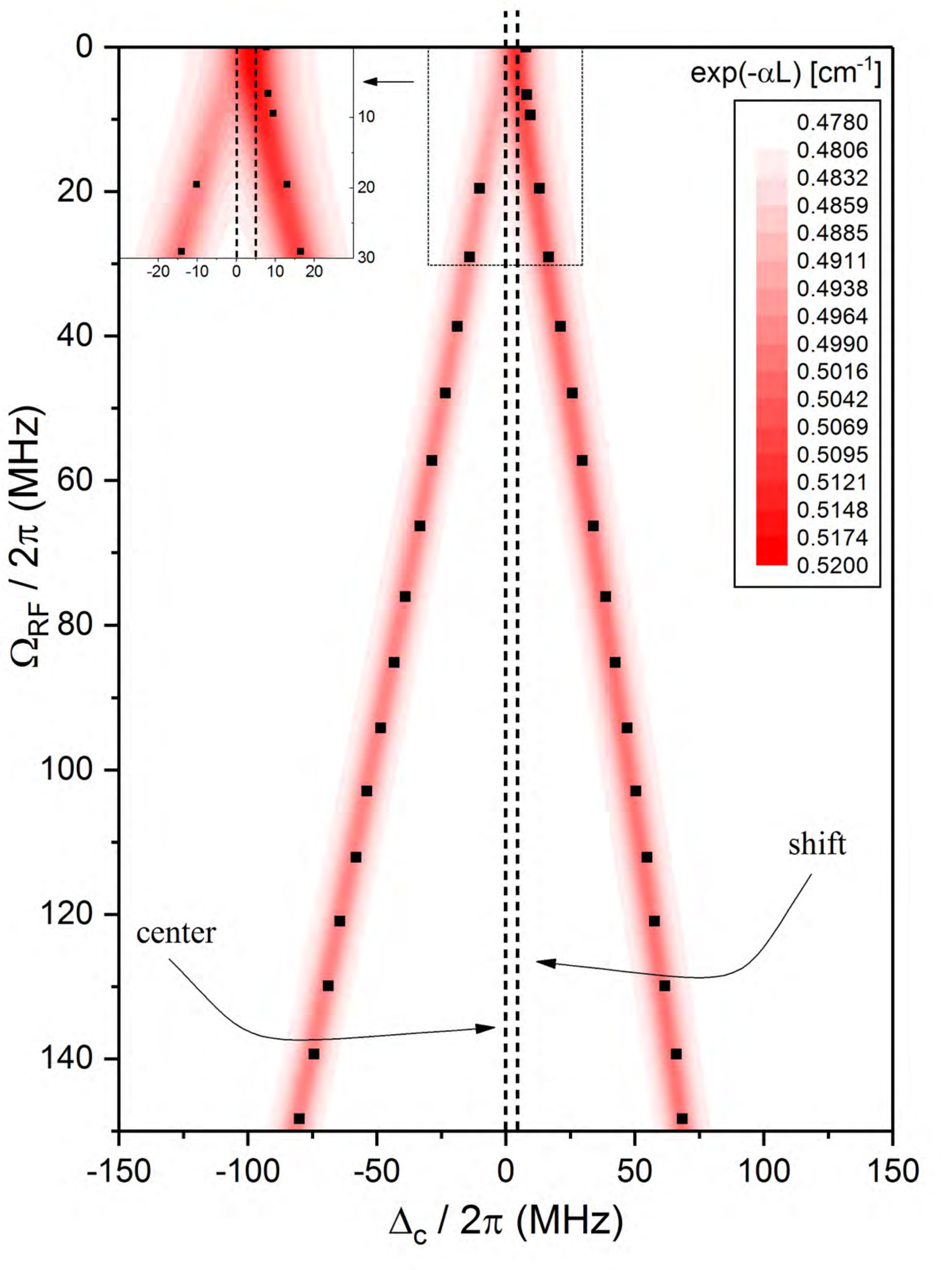}}\\
\vspace*{-2mm}
{\hspace{-1mm}\tiny{(a) \hspace{77mm} (d)}}\\
\vspace*{2mm}
\scalebox{.27}{\includegraphics{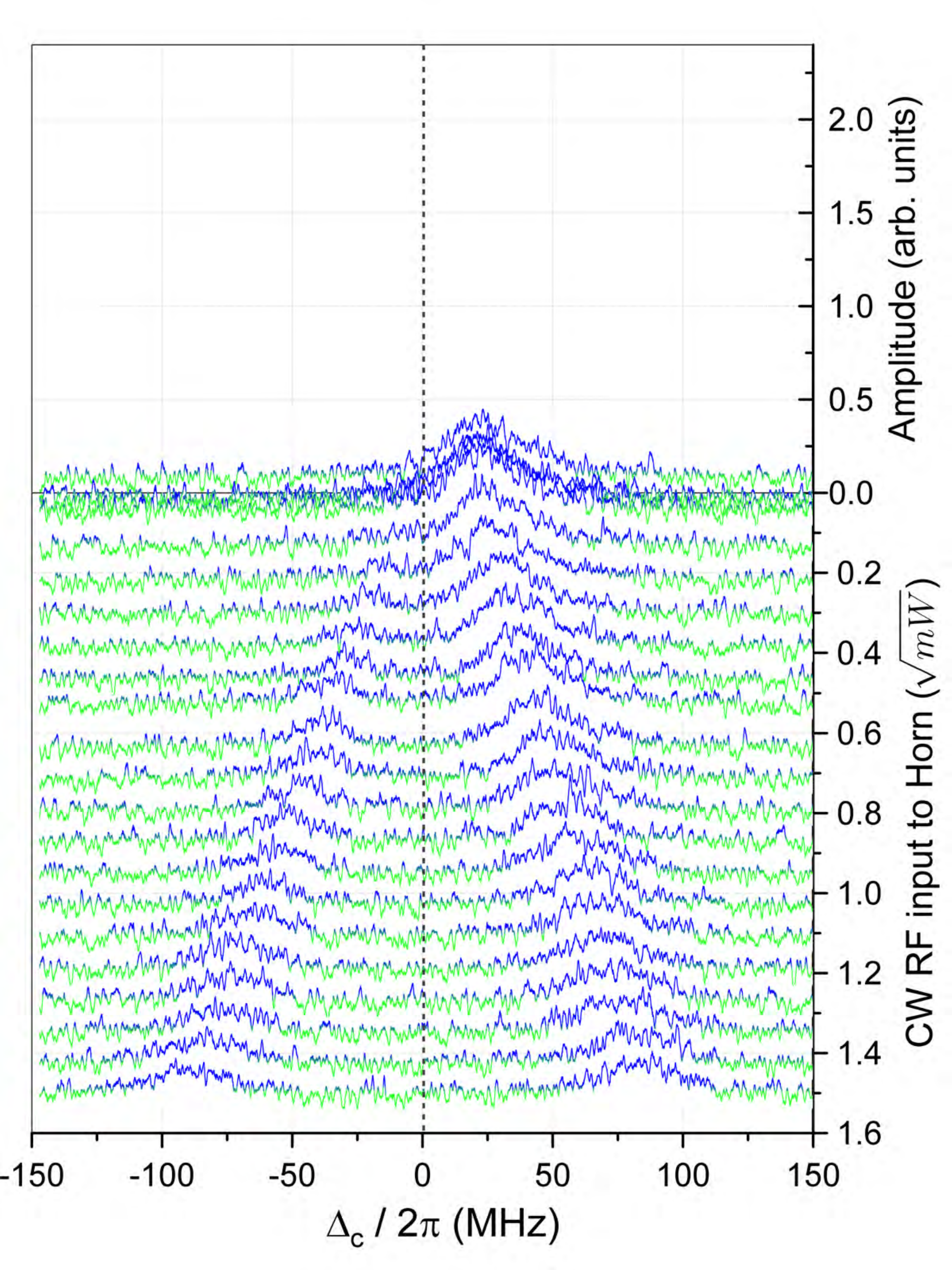}} \hspace{20mm}
\scalebox{.27}{\includegraphics{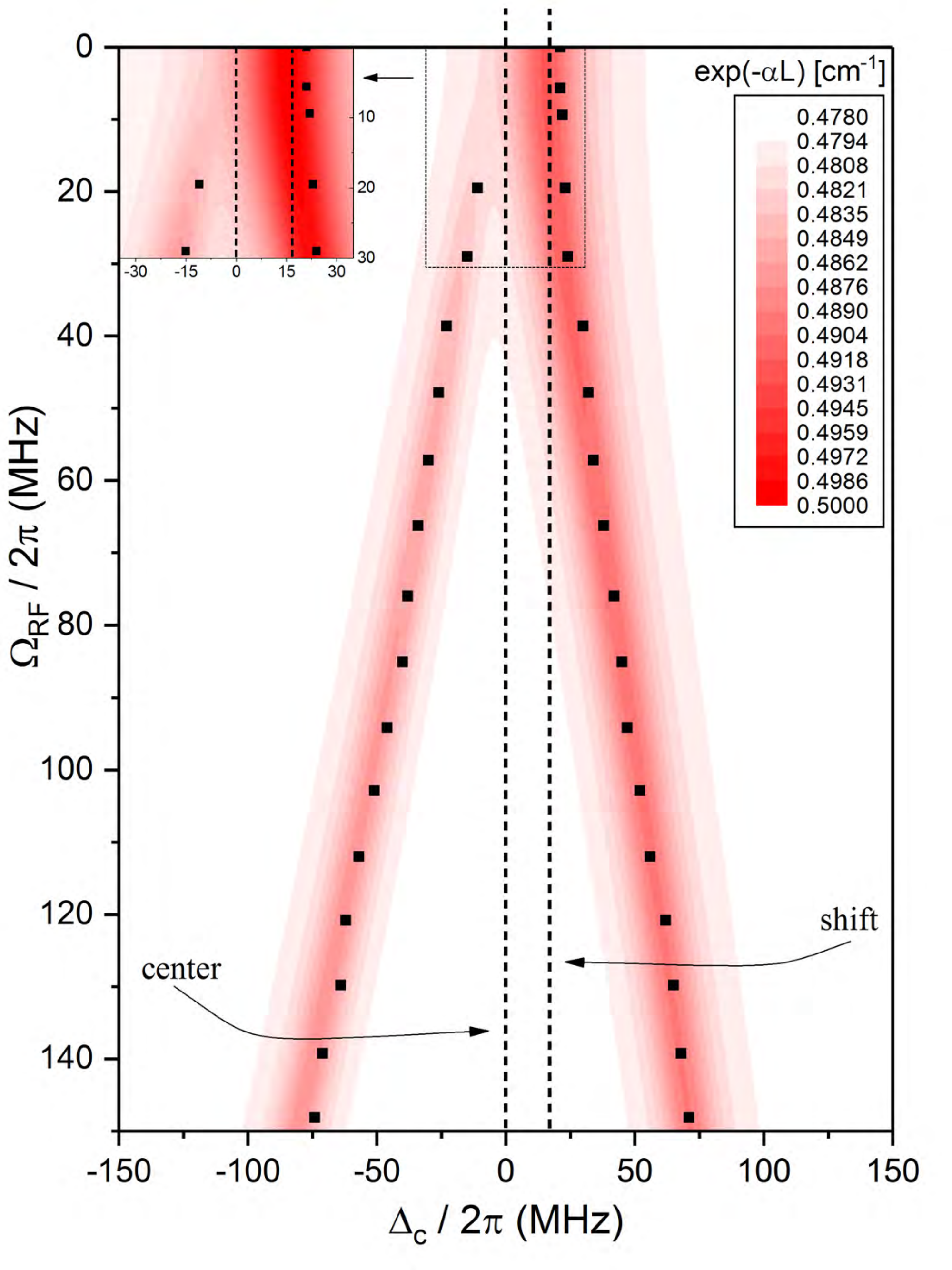}}\\
\vspace*{-2mm}
{\hspace{-1mm}\tiny{(b) \hspace{77mm} (e)}}\\
\vspace*{2mm}
\scalebox{.27}{\includegraphics{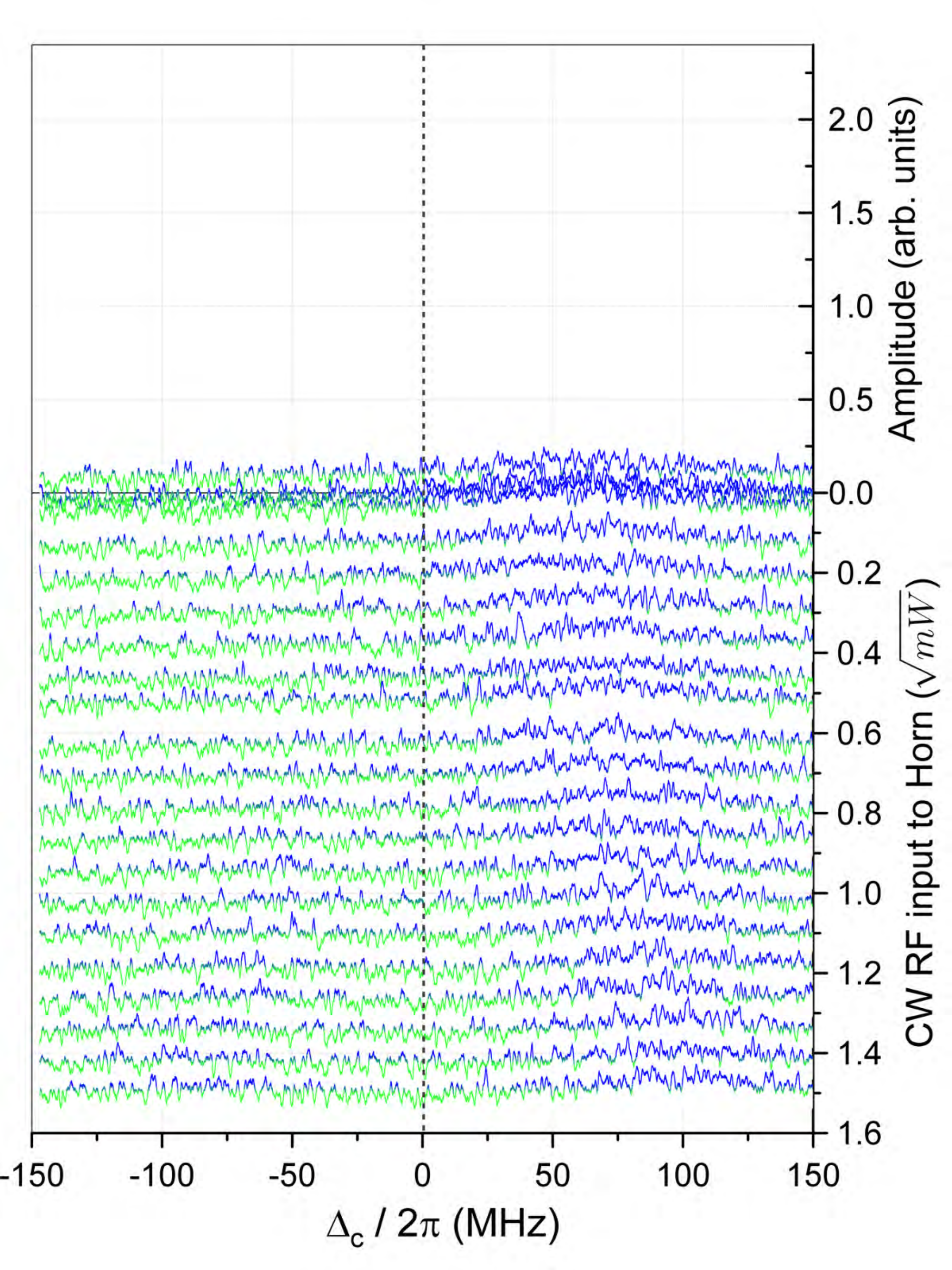}} \hspace{20mm}
\scalebox{.27}{\includegraphics{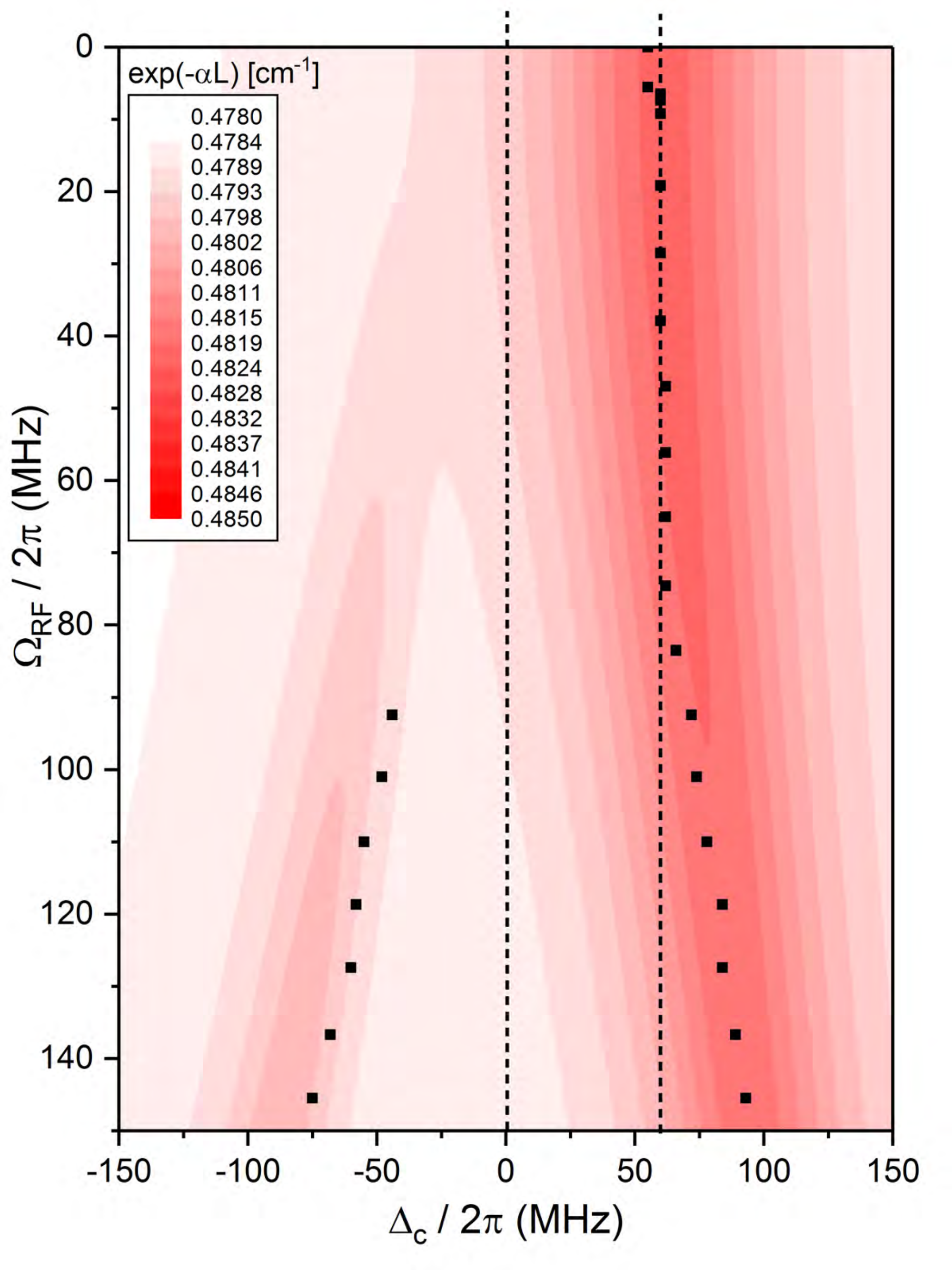}}\\
\vspace*{-2mm}
{\hspace{-1mm}\tiny{(c) \hspace{77mm} (f)}}\\
\caption{Experimental data (a) - (c) and model results (d) - (f) for Filter 1.  The EIT signal is plotted as a function of coupling laser detuning $\Delta_c$ for different RF E-field strengths and different noise sources. This dataset is for
a coherent resonant RF of 19.7825~GHz and corresponds to the following $^{85}$Rb 4-level atomic system,  5$S_{1/2}$-5$P_{3/2}$-57$S_{1/2}$-57$P_{1/2}$. (a)/(d) -12~dB attenuation, (b)/(e) -6~dB attenuation, (c)/(f) 0~dB attenuation. The squares on the plots shown in (d), (e), and (f) correspond to the peaks of the experimental EIT data shown in (a), (b), and (c). The dashed line at $\Delta_c=0$ is used as reference in order to guide the eye to show shifts from $\Delta_c=0$.}
\label{noiseEIT1}
\end{figure*}

\begin{figure*}
\centering
\scalebox{.27}{\includegraphics{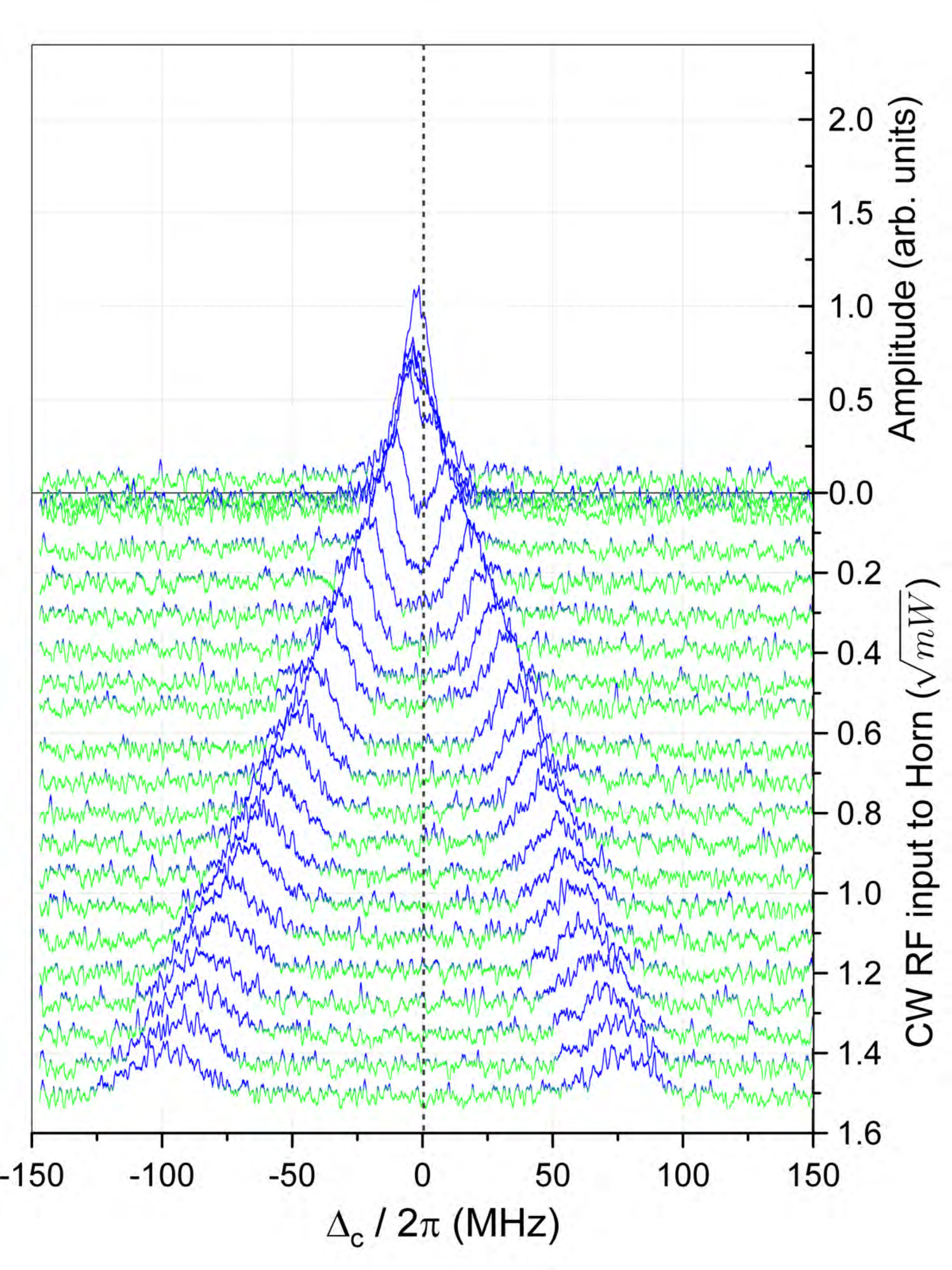}} \hspace{20mm}
\scalebox{.27}{\includegraphics{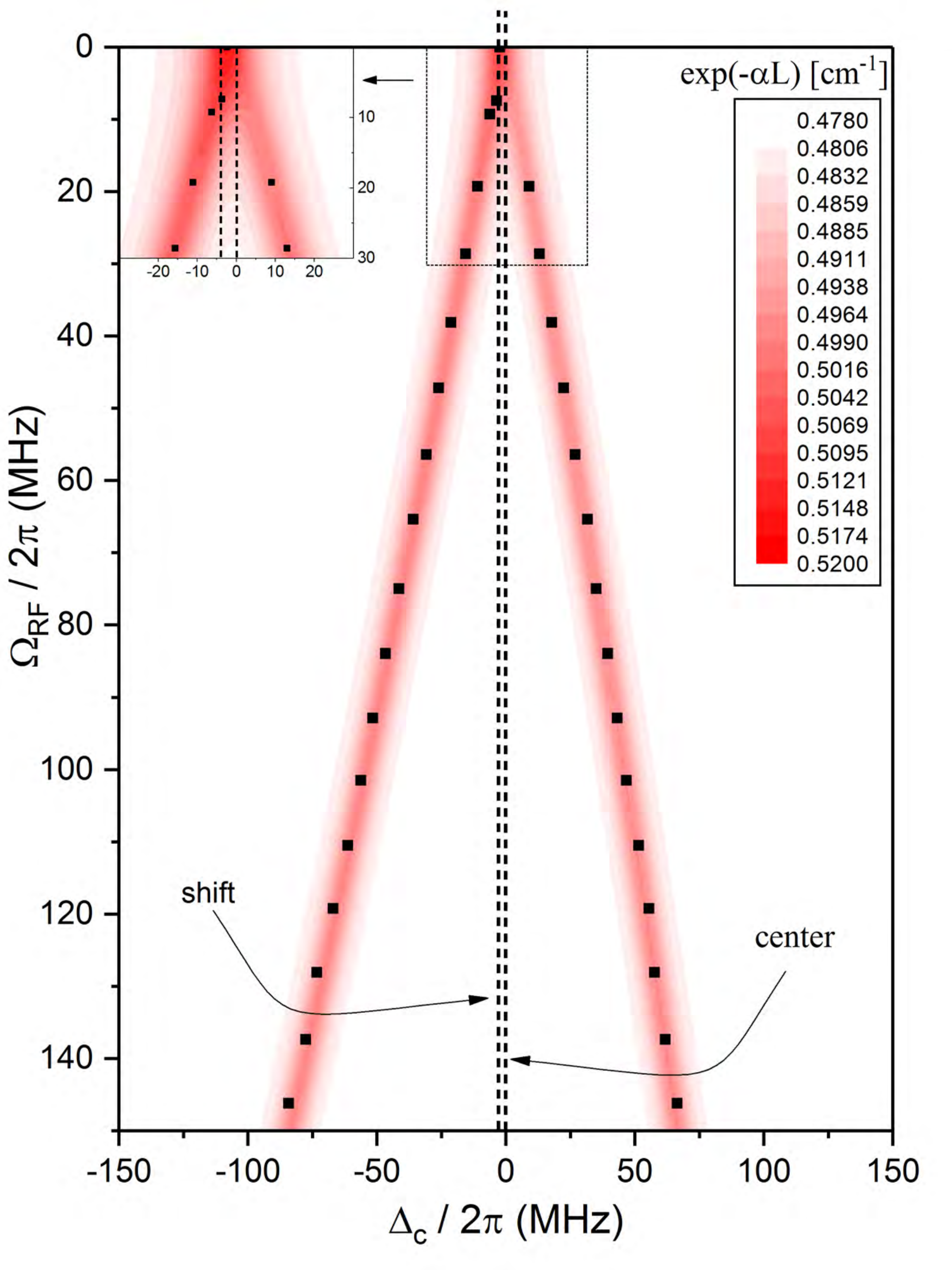}}\\
\vspace*{-2mm}
{\hspace{-1mm}\tiny{(a) \hspace{77mm} (d)}}\\
\vspace*{2mm}
\scalebox{.27}{\includegraphics{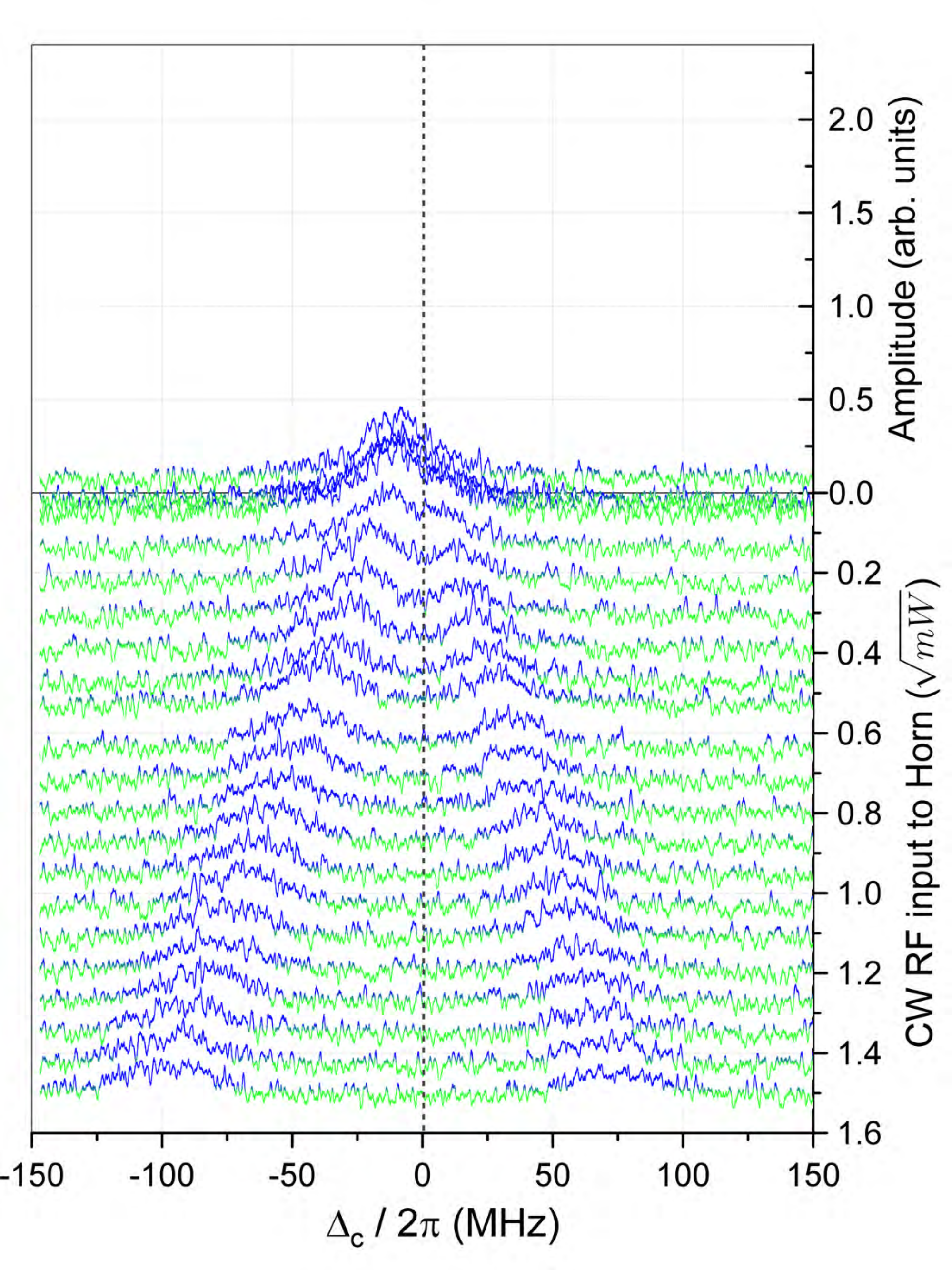}} \hspace{20mm}
\scalebox{.27}{\includegraphics{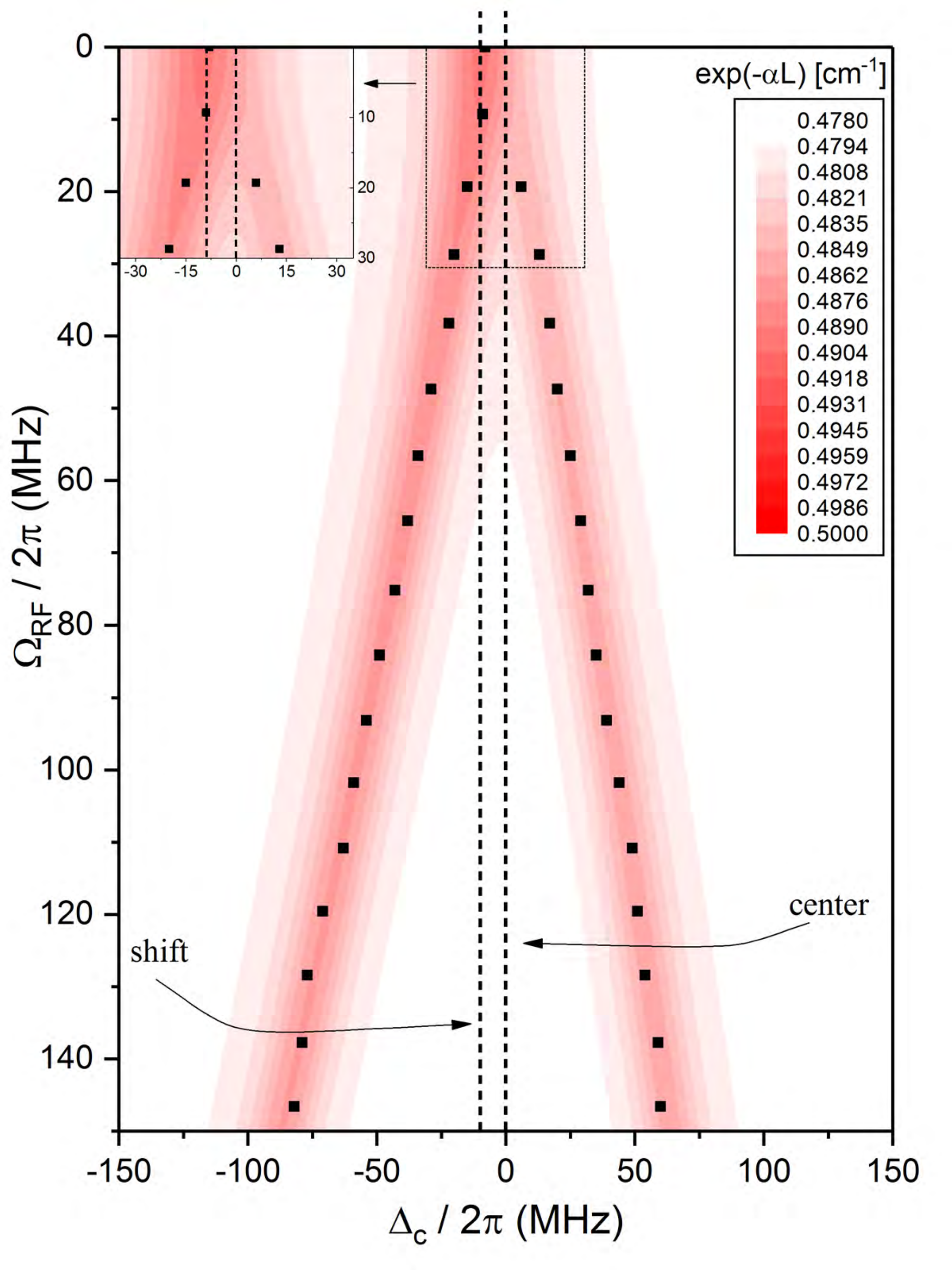}}\\
\vspace*{-2mm}
{\hspace{-1mm}\tiny{(b) \hspace{77mm} (e)}}\\
\vspace*{2mm}
\scalebox{.27}{\includegraphics{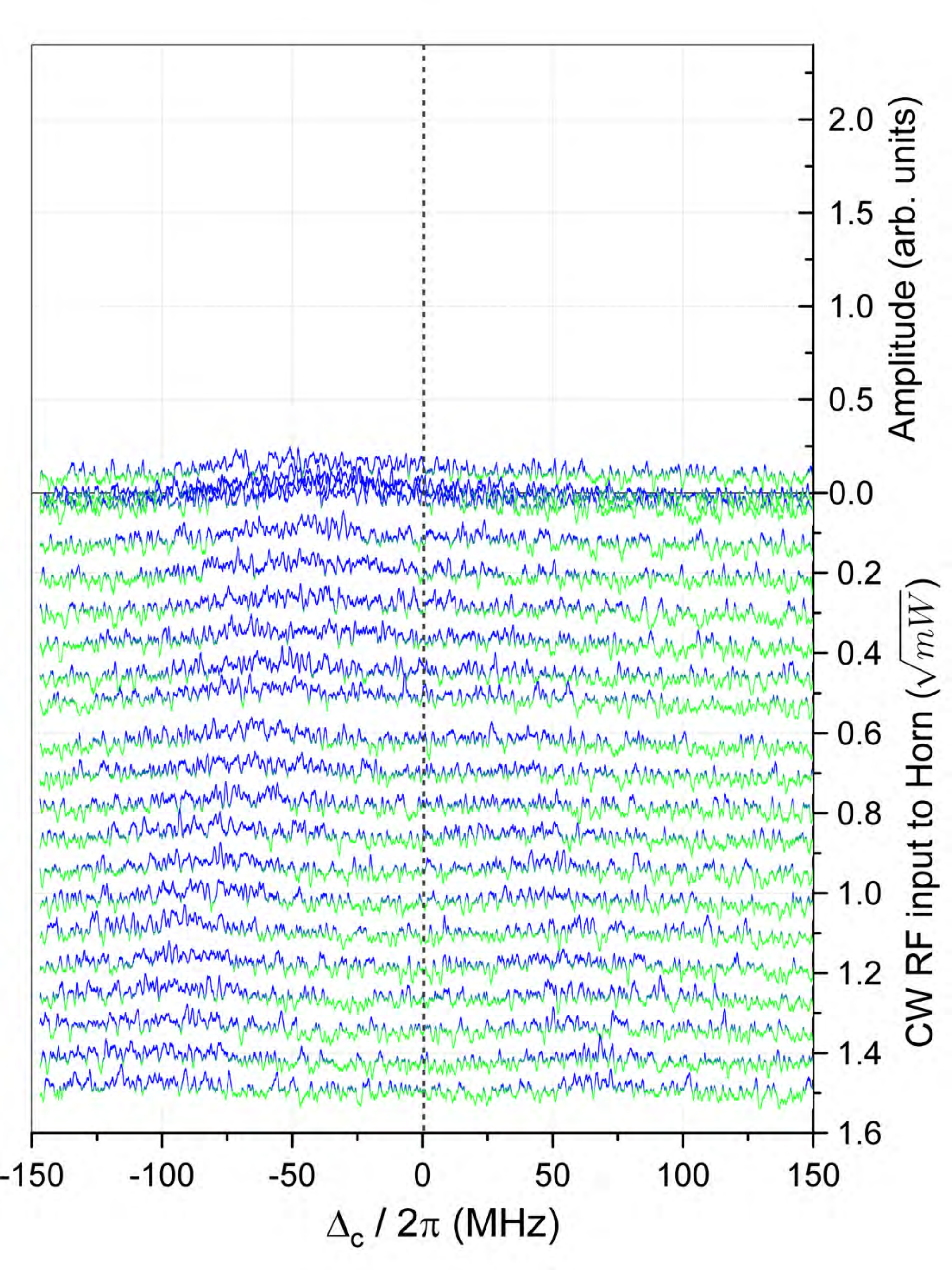}} \hspace{20mm}
\scalebox{.27}{\includegraphics{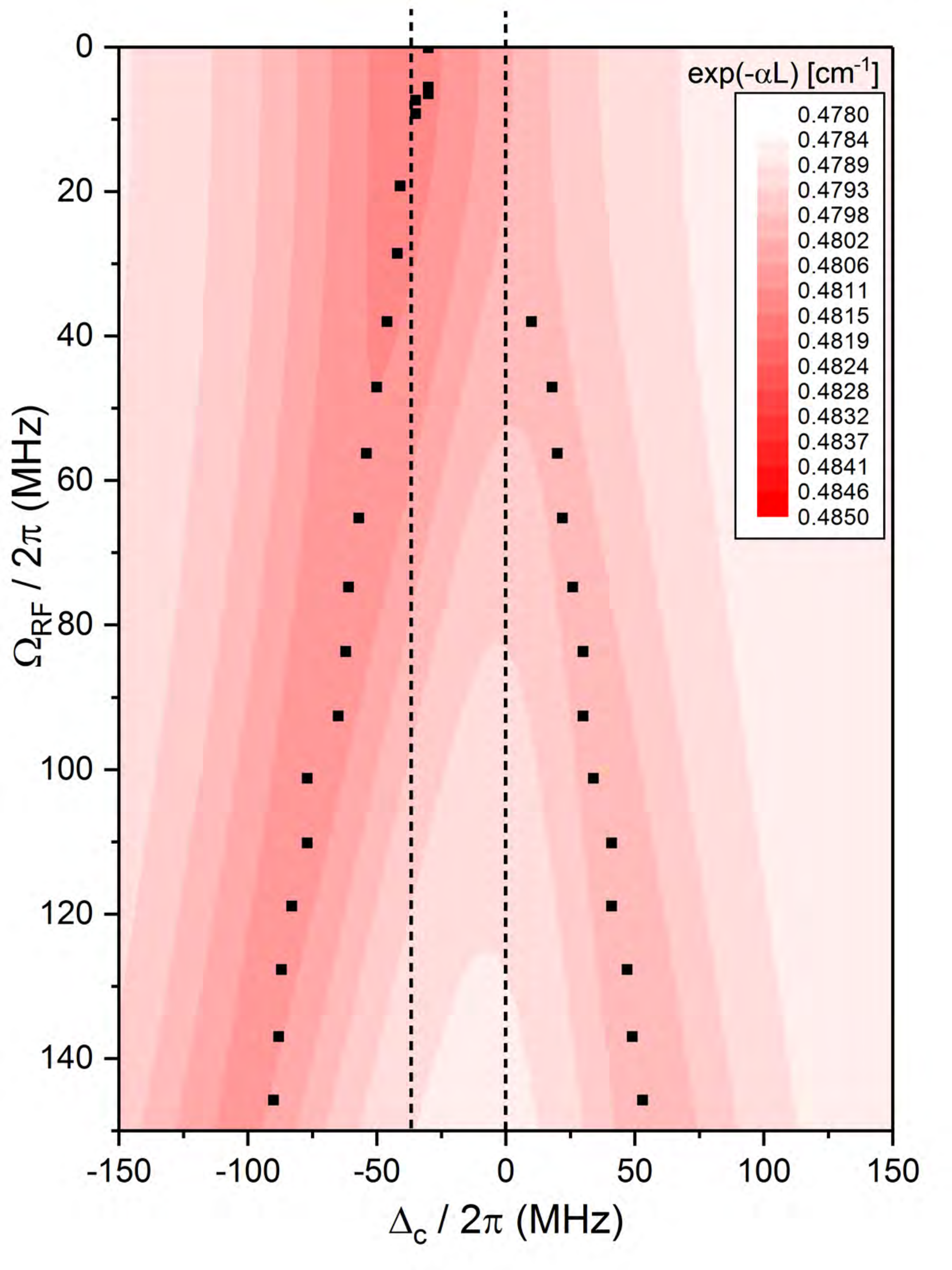}}\\
\vspace*{-2mm}
{\hspace{-1mm}\tiny{(c) \hspace{77mm} (f)}}\\
\caption{Experimental data (a) - (c) and model results (d) - (f) for Filter 2. The squares on the plots shown in (d), (e), and (f) correspond to the peaks of the experimental EIT data shown in (a), (b), and (c). Additional details are the same as in Fig.~\ref{noiseEIT1}.}
\label{noiseEIT2}
\end{figure*}

\begin{figure*}
	\centering
	\scalebox{.27}{\includegraphics{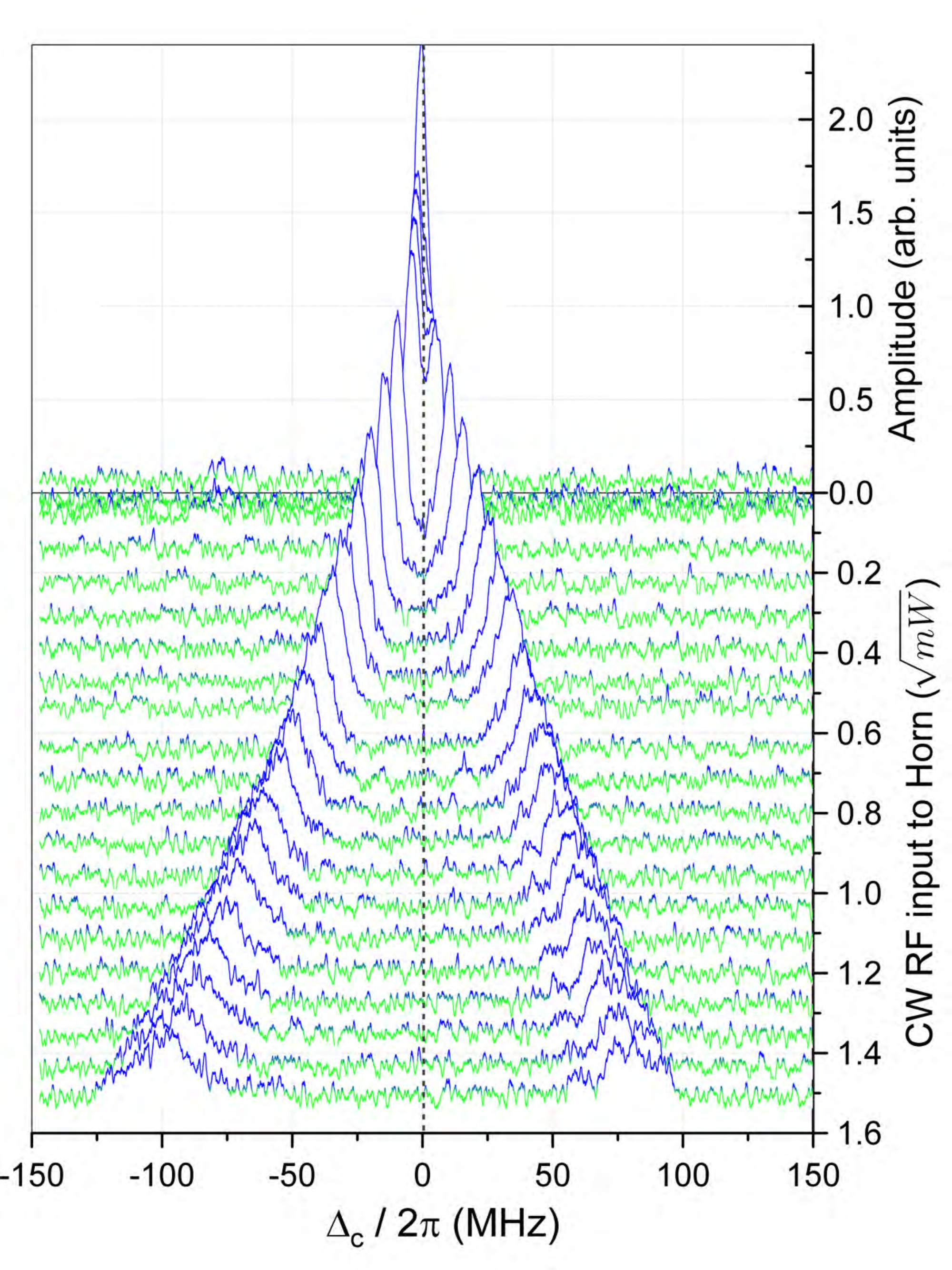}} \hspace{20mm}
	\scalebox{.27}{\includegraphics{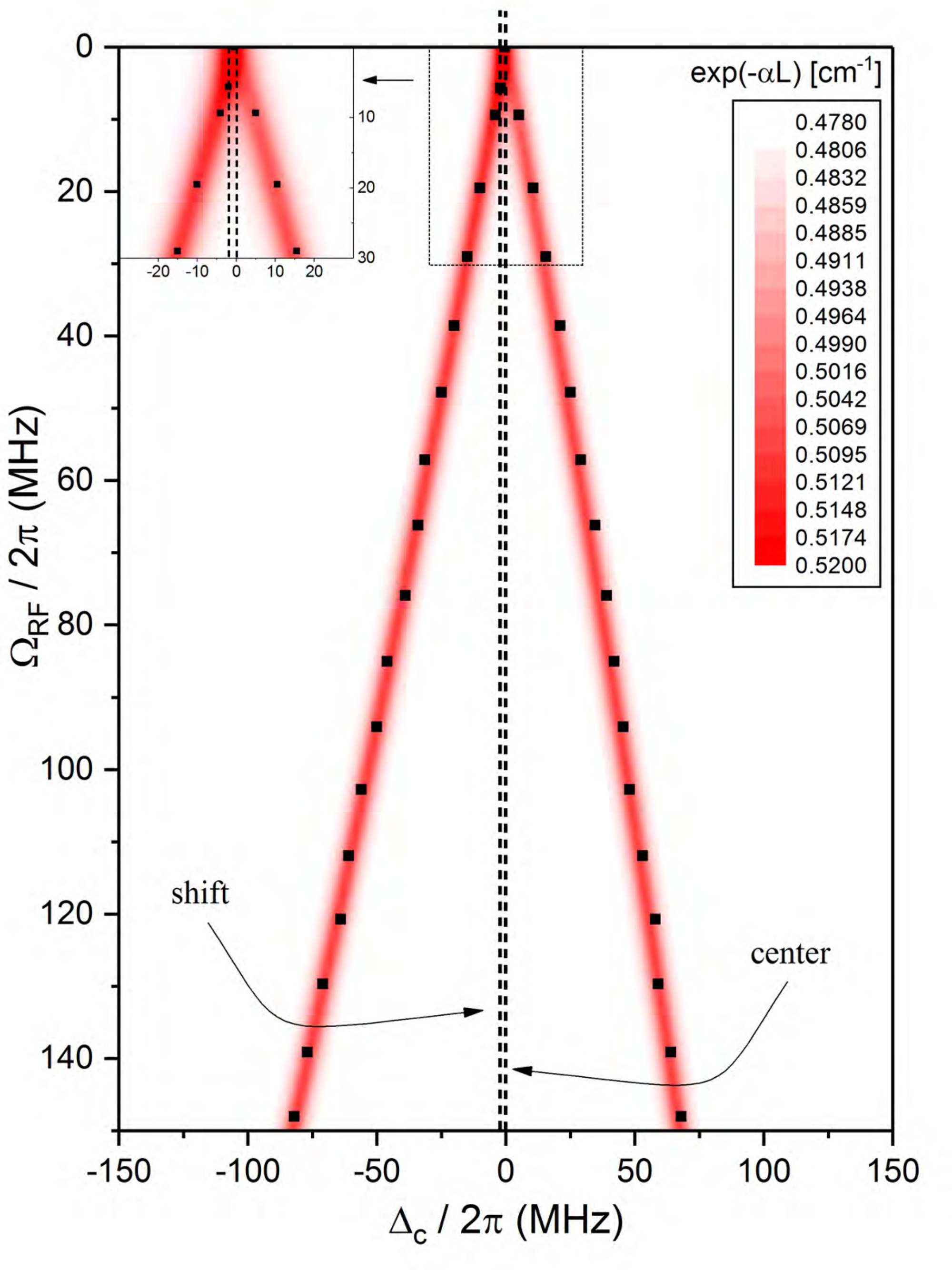}}\\
	\vspace*{-2mm}
	{\hspace{-1mm}\tiny{(a) \hspace{77mm} (d)}}\\
	\vspace*{2mm}
	\scalebox{.27}{\includegraphics{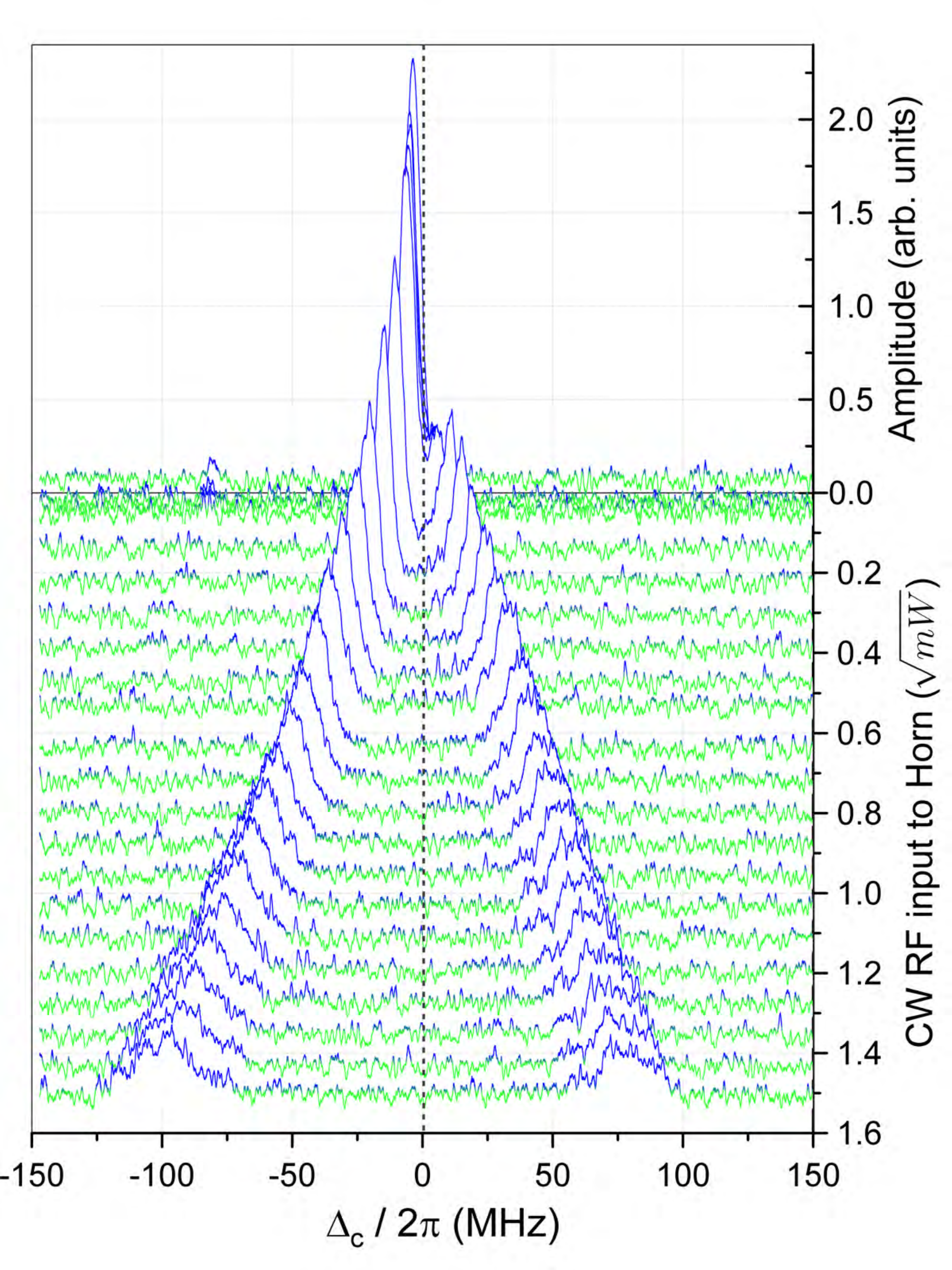}} \hspace{20mm}
	\scalebox{.27}{\includegraphics{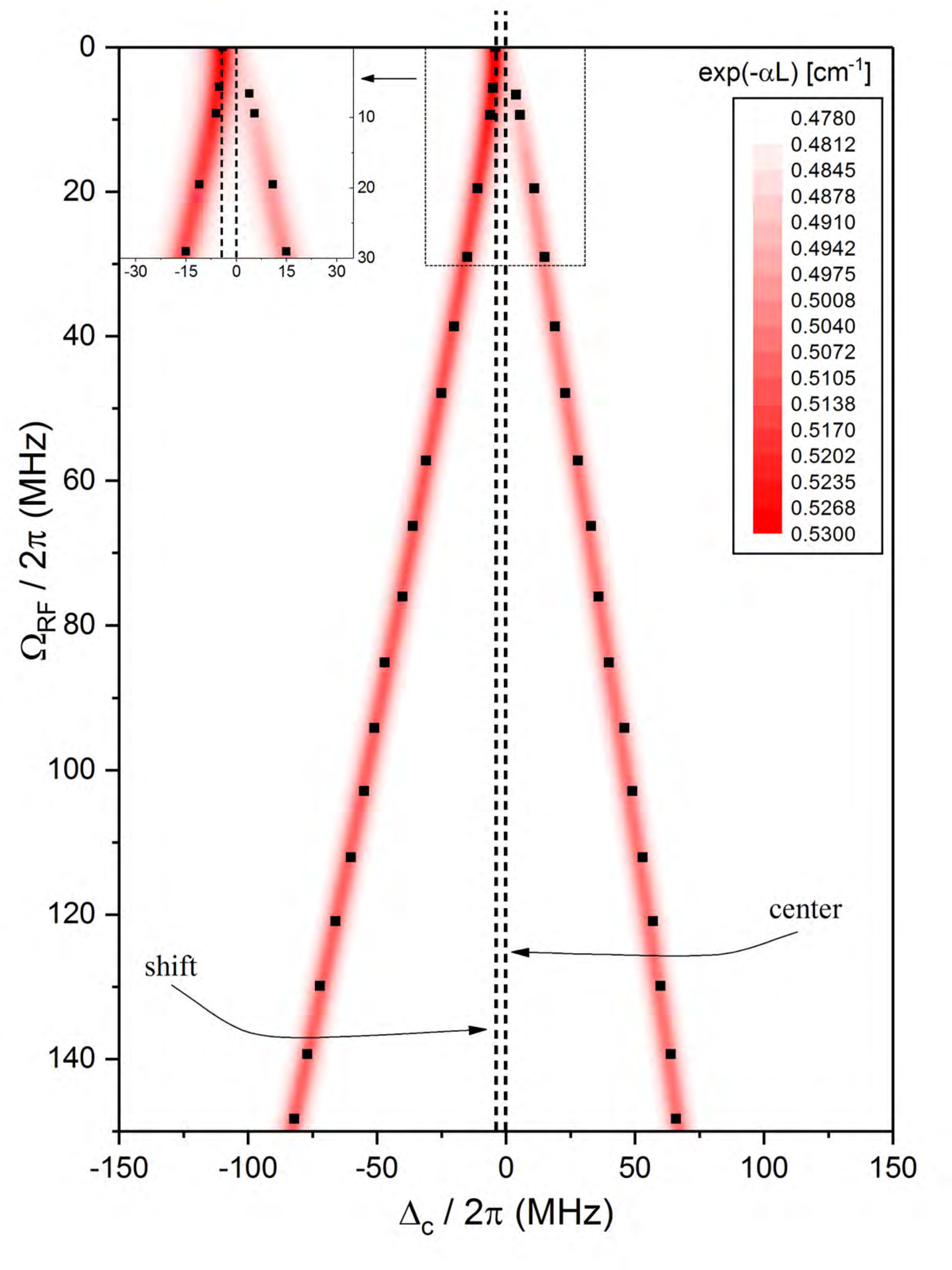}}\\
	\vspace*{-2mm}
	{\hspace{-1mm}\tiny{(b) \hspace{77mm} (e)}}\\
	\vspace*{2mm}
	\scalebox{.27}{\includegraphics{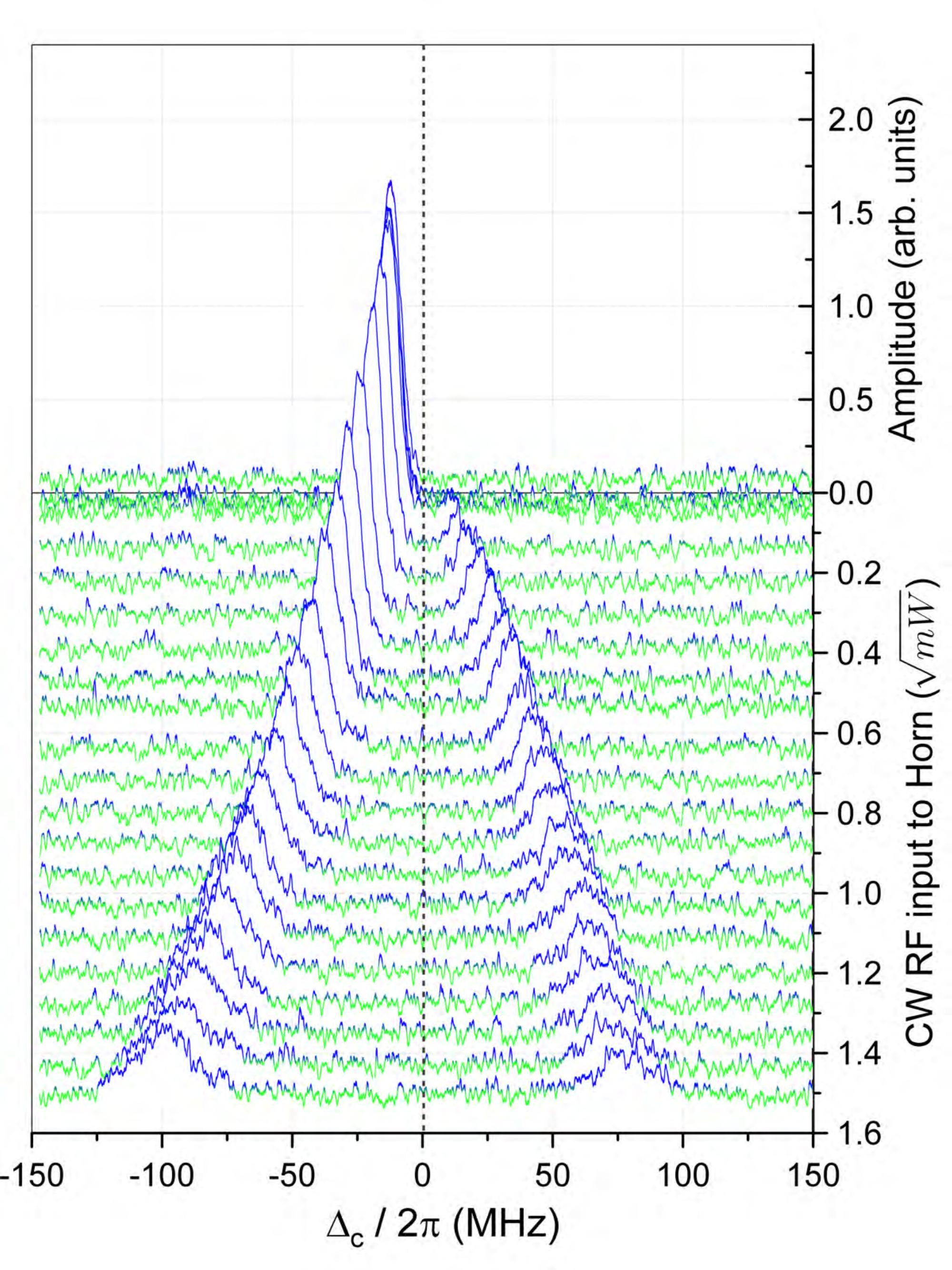}} \hspace{20mm}
	\scalebox{.27}{\includegraphics{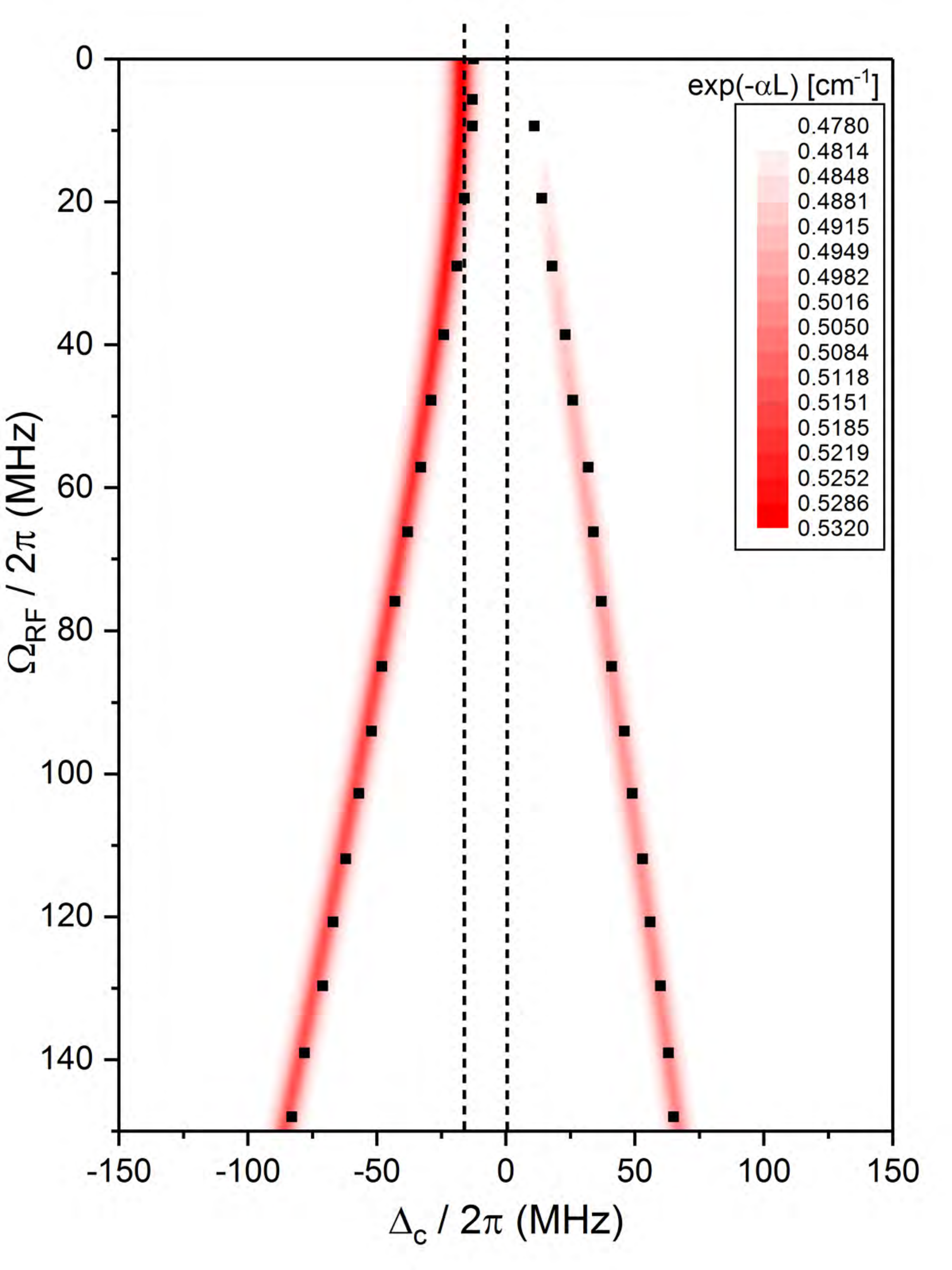}}\\
	\vspace*{-2mm}
	{\hspace{-1mm}\tiny{(c) \hspace{77mm} (f)}}\\
	\caption{Experimental data (a) - (c) and model results (d) - (f) for Filter 3. The squares on the plots shown in (d), (e), and (f) correspond to the peaks of the experimental EIT data shown in (a), (b), and (c). Additional details are the same as in Fig.~\ref{noiseEIT1}.}
	\label{noiseEIT3}
\end{figure*}

\begin{figure*}
	\centering
	\scalebox{.27}{\includegraphics{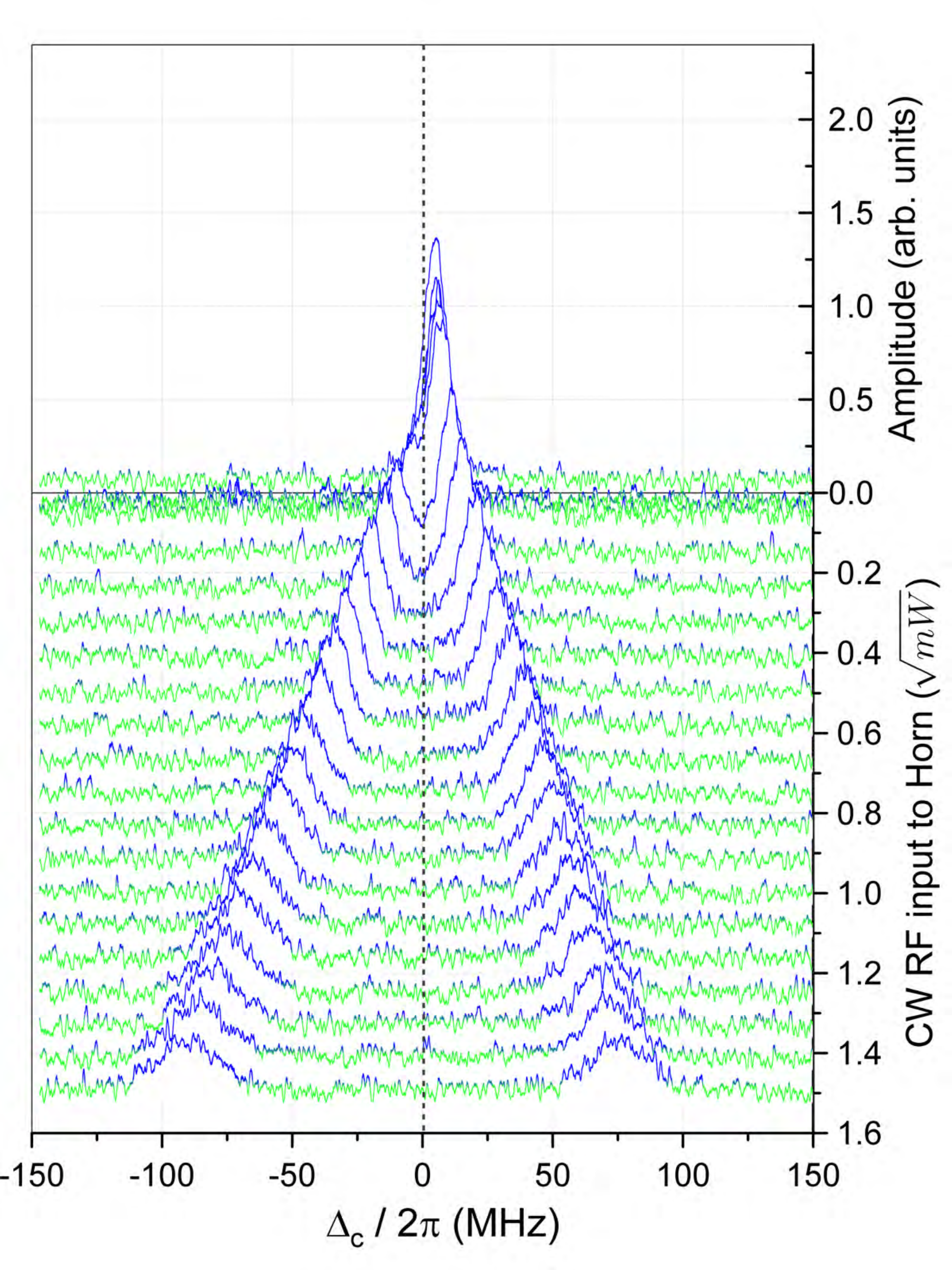}} \hspace{20mm}
	\scalebox{.27}{\includegraphics{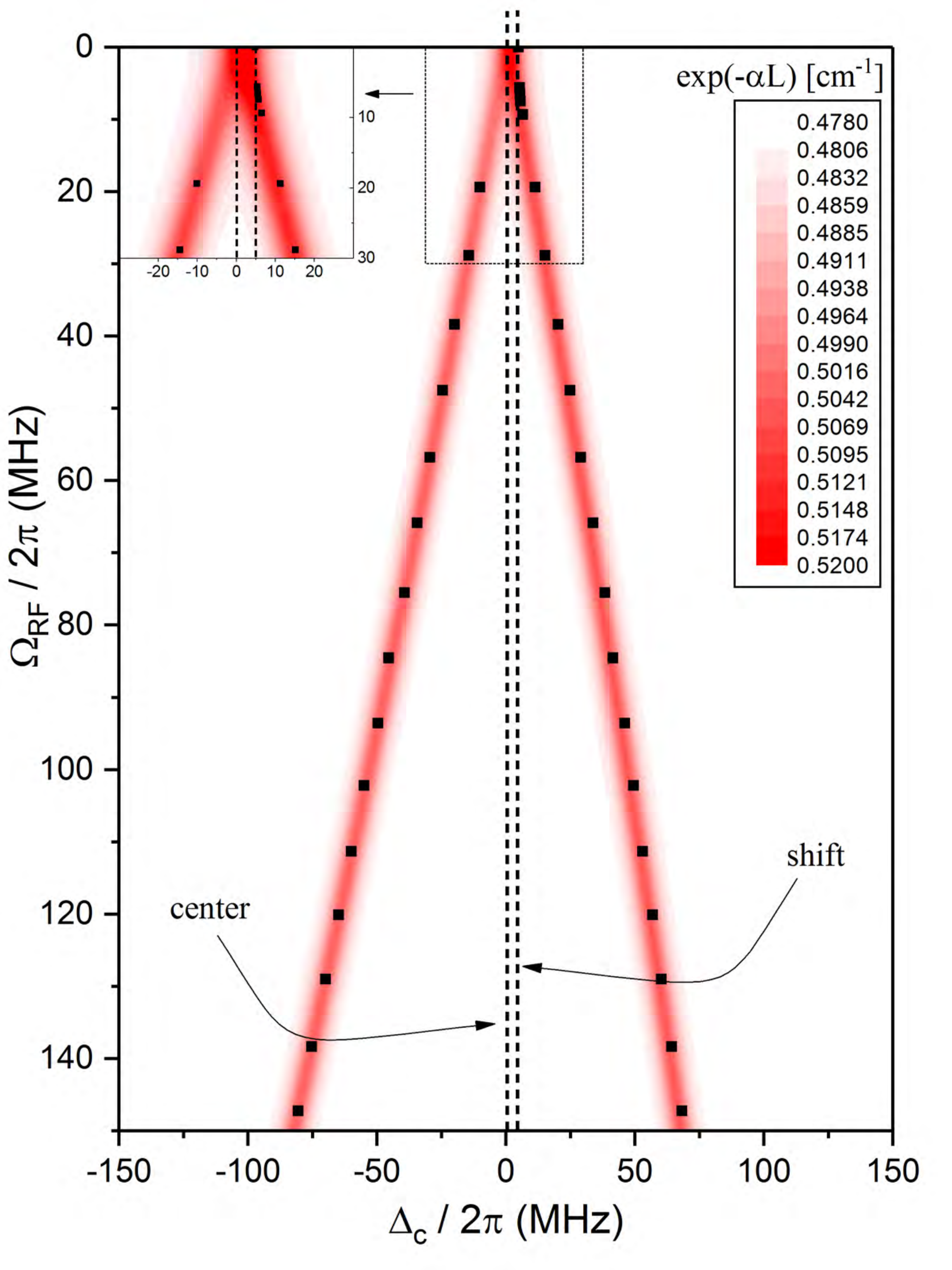}}\\
	\vspace*{-2mm}
	{\hspace{-1mm}\tiny{(a) \hspace{77mm} (d)}}\\
	\vspace*{2mm}
	\scalebox{.27}{\includegraphics{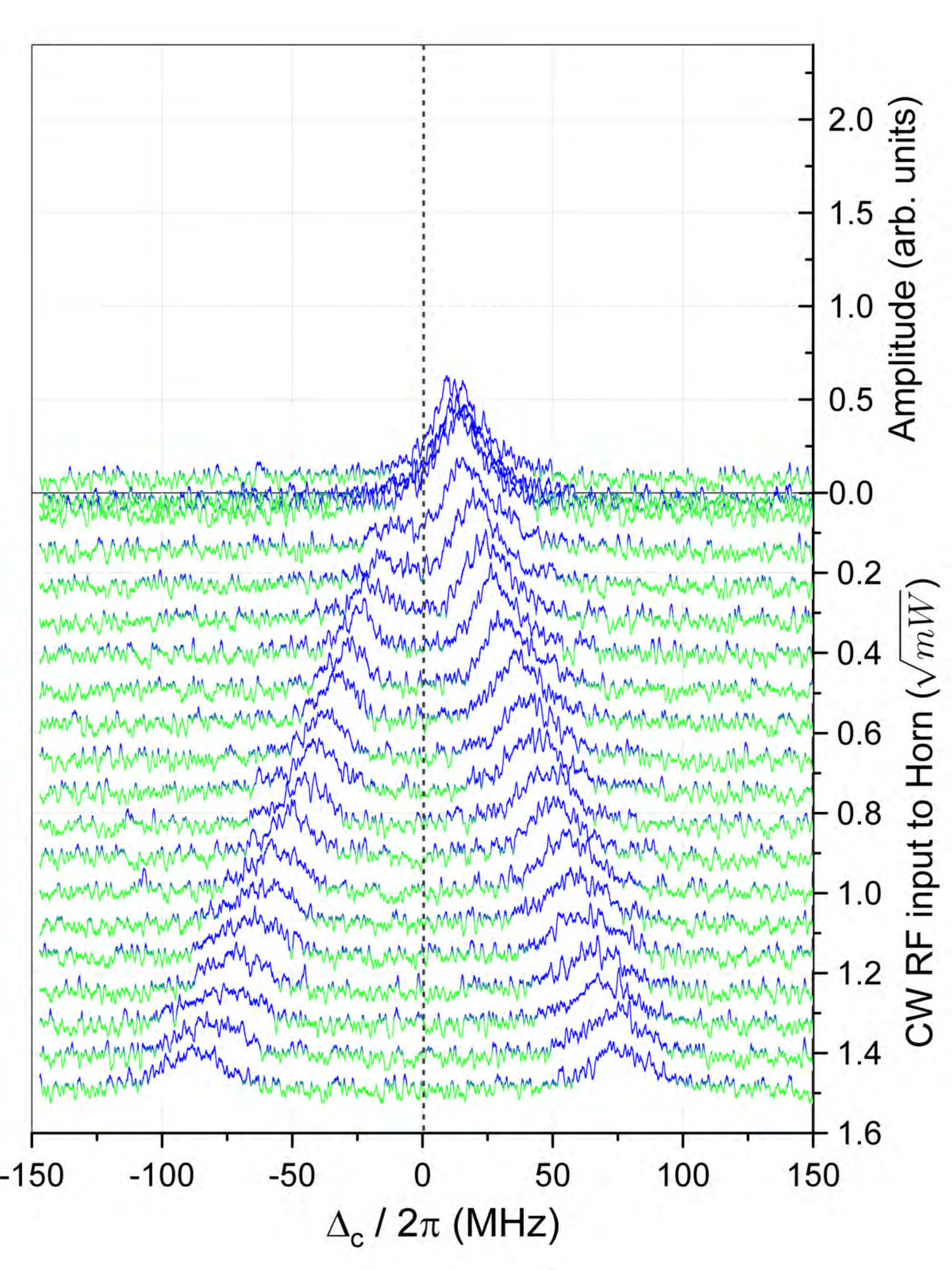}} \hspace{20mm}
	\scalebox{.27}{\includegraphics{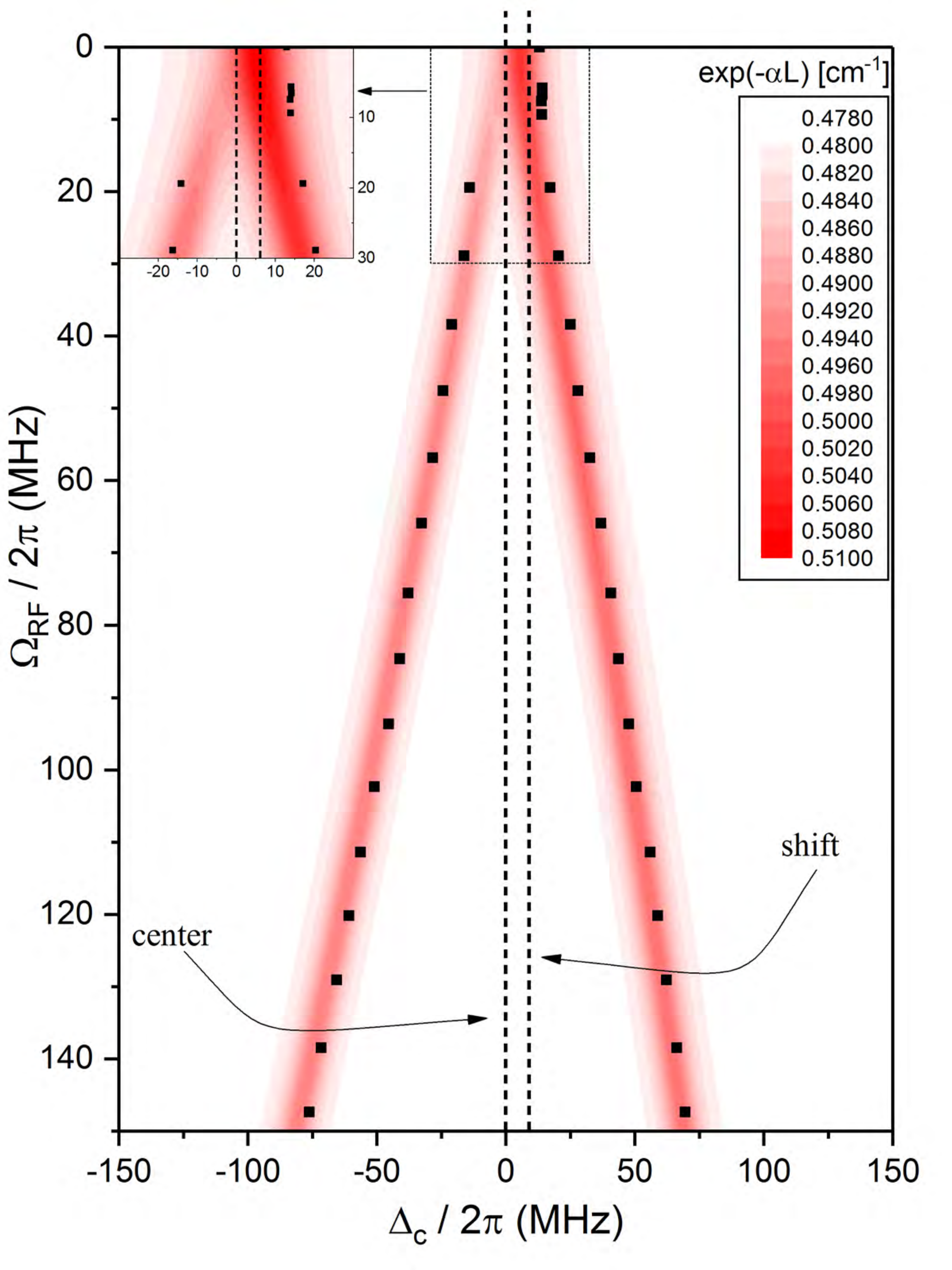}}\\
	\vspace*{-2mm}
	{\hspace{-1mm}\tiny{(b) \hspace{77mm} (e)}}\\
	\vspace*{2mm}
	\scalebox{.27}{\includegraphics{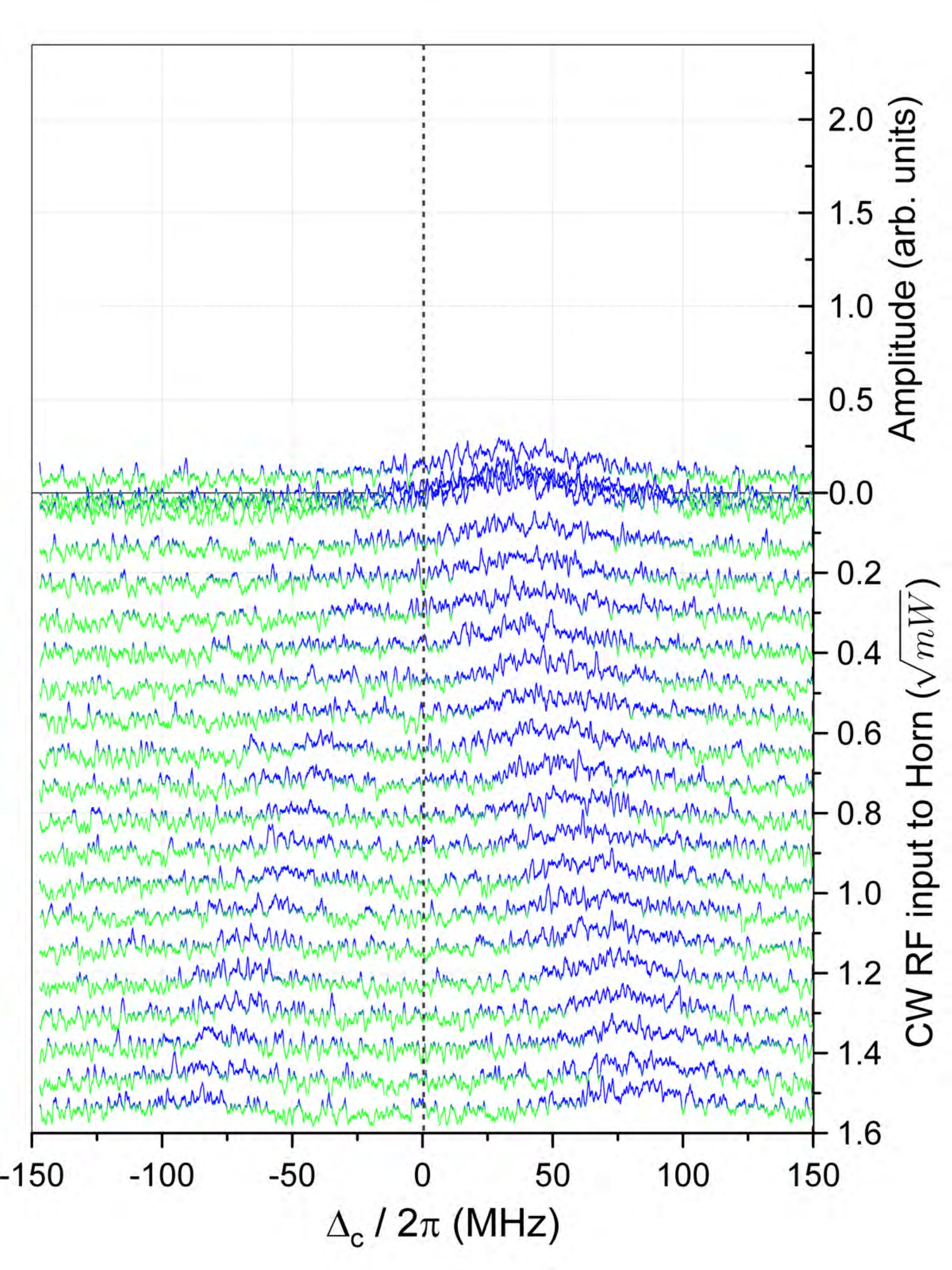}} \hspace{20mm}
	\scalebox{.27}{\includegraphics{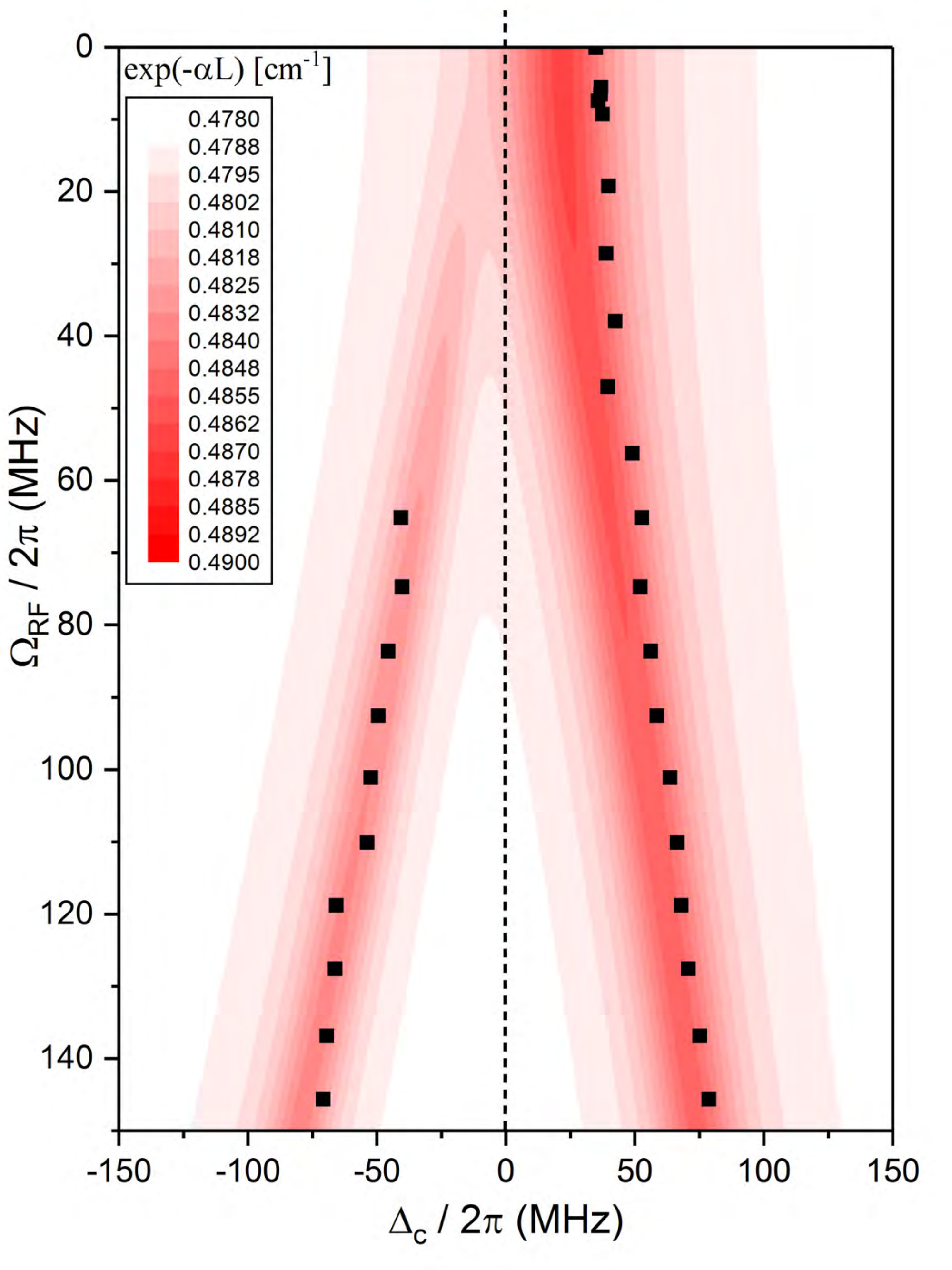}}\\
	\vspace*{-2mm}
	{\hspace{-1mm}\tiny{(c) \hspace{77mm} (f)}}\\
	\caption{Experimental data (a) - (c) and model results (d) - (f) for Filter combination 1/3. The squares on the plots shown in (d), (e), and (f) correspond to the peaks of the experimental EIT data shown in (a), (b), and (c). Additional details are the same as in Fig.~\ref{noiseEIT1}.}
	\label{noiseEIT13}
\end{figure*}

\begin{figure}
\centering
\scalebox{.32}{\includegraphics{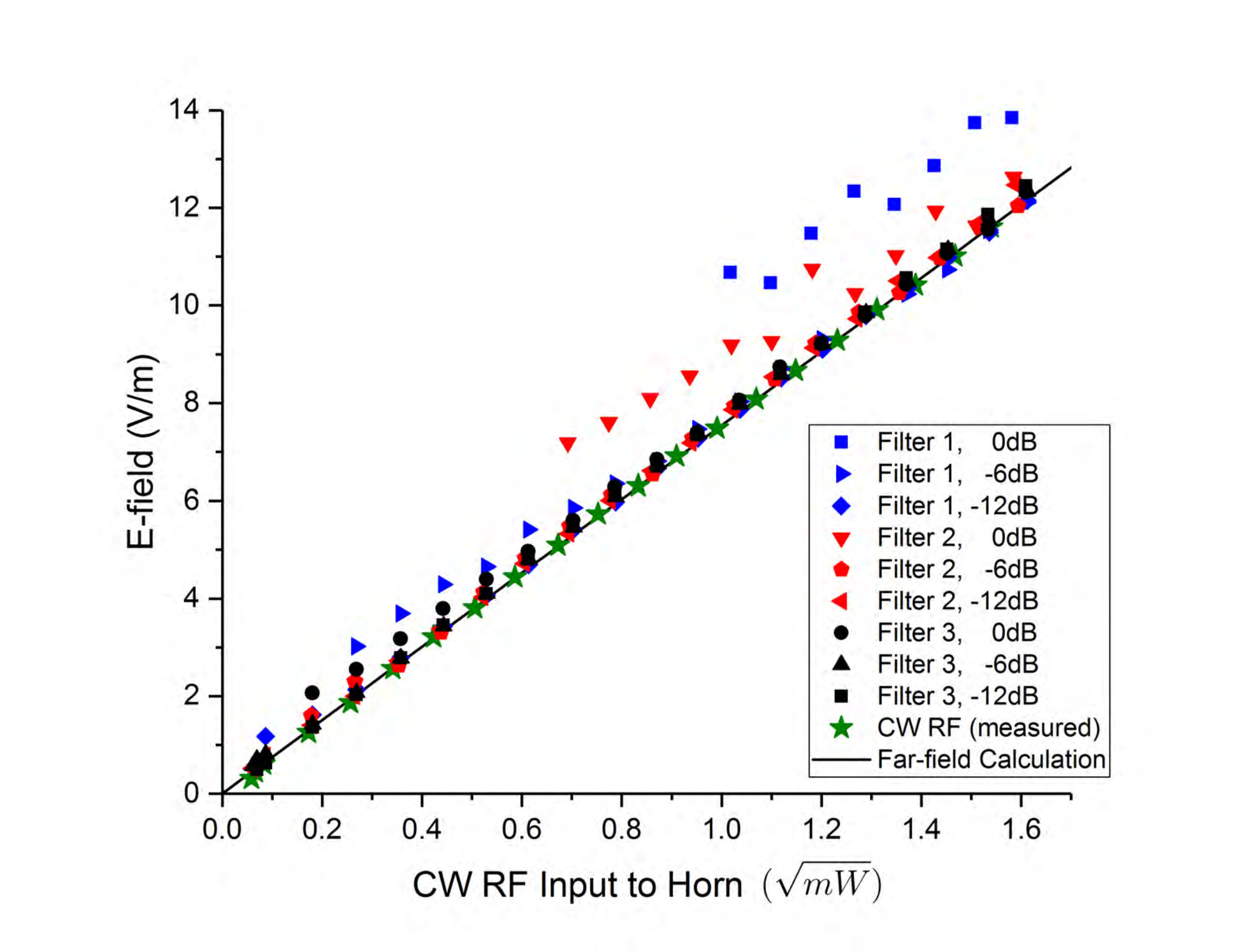}}\\
\caption{Calculated E-field strength from eq.~(\ref{mage}) as a function input CW RF power to the antenna for various noise conditions (Filter \#, integrated noise power). We use $\wp=1120 e a_0$ (radial part of $3360 e a_0$ and an angular part of $1/3$.)}
\label{Efields}
\end{figure}

\begin{figure}
	\centering
	\scalebox{.32}{\includegraphics{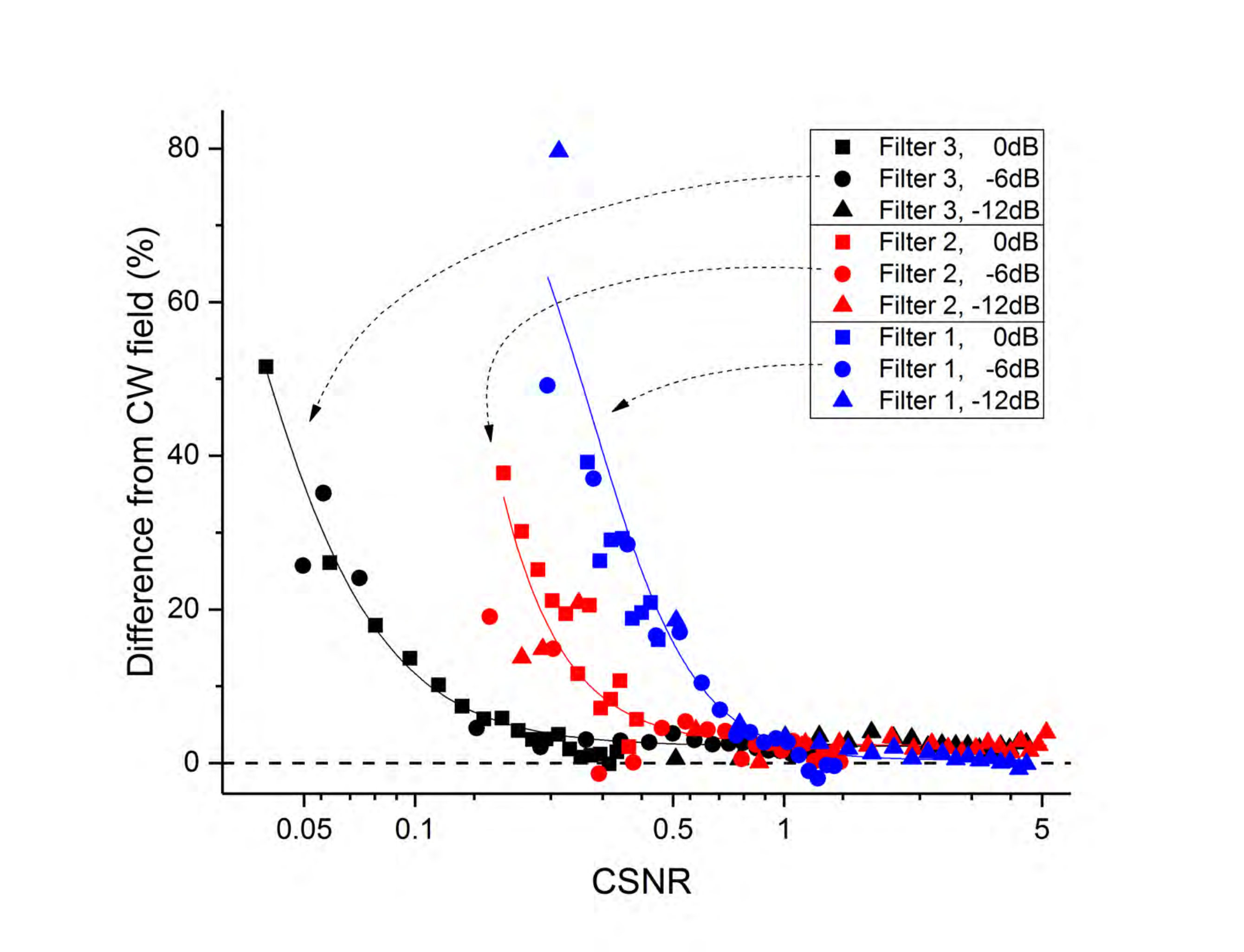}}\\
	\caption{Percent difference between E-field measurements with added noise and measurements with no noise [based on eq.~(\ref{mage})], plotted vs. the coherent-signal-to-noise-power ratio (CSNR).}
	\label{Error_SNR}
\end{figure}

Fig.~\ref{Efields} shows the measured E-fields obtained from the EIT signals from all three noise filters (i.e., the results in Fig.~\ref{noiseEIT1}-\ref{noiseEIT3}) using eq.~(\ref{mage}). In these calculations we use $\wp=1120 e a_0$ (where $e$ is the elementary charge; $a_0=0.529177\times 10^{-10}$~m and is the Bohr radius).  This dipole moment for the resonant RF transition is composed of a radial part of $3360 e a_0$ and an angular part of $1/3$. The black line is a far-field calculation of the expected \ef{} strength (taking into account the $A_{sw}$ enhancement factor) and agrees with the measured E-field for the CW RF no-noise case. Adding noise causes an increase in the measured \ef{}, by an amount that depends on the frequency band of the noise, the noise power, and the strength of the CW field under test. For most noise conditions, the measured E-field strengths with noise are larger by some amount when compared to the no-noise case.  The highest noise levels (i.e., 0~dB) with Filters 1 and 2 does show a dramatic increase in the AT-splitting and the apparent E-field strength. In actuality, the E-field strength of the coherent signal in the vapor cell has not increased at all. It is only the AT-splitting that has increased due to the noise. This point is further explained by the model presented in the next section.

To better illustrate the effect of the noise, Fig.~\ref{Error_SNR} shows the percent difference of the apparent E-field obtained using eq.~(\ref{mage}) from the no-noise case versus the coherent-signal-to-noise-power ratio (CSNR). The CSNR is calculated using the CW RF power and the integrated noise power at the input to the horn antenna. We see that above a CSNR of 1, the percent difference approaches zero. Below a CSNR of 1 the effect on the \efm{} strongly depends on the frequency band of the noise. When the noise is red-shifted compared to the RF transition frequency (i.e., Filter 3), the noise primarily broadens the EIT signal, and has less effect on shifting the EIT peaks compared to other filters.  This will be shown and discussed in more detail in the next session. Thus, when the noise is red-shifted relative to the RF transition, it has a minimal effect on the ability to measure CW RF \efs{}. Even with the strongest noise power (black squares in Fig.~\ref{Error_SNR}), the \efm{} is only strongly affected below a CSNR of 0.2. The blue-shifted noise (i.e., Filter 1) has the strongest effect on the \efms{}. In the Filter 1 case, the \efms{} significantly differ from the CW RF below a CSNR of 1 and differ by $\sim 80\%$ below a CSNR of 0.5. The influence of BLWGN on E-fields measurement for the on-resonance case (Filter 2) falls between the red-shifted and blue-shift cases (Filter 1 and Filter 3).

The experimental results in Figs.~\ref{EIT} and \ref{noiseEIT1}-\ref{Error_SNR} serve to illustrate the typical effects of BWLGN on Rydberg-EIT-AT spectra. The baseline measurement in Fig.~\ref{EIT} shows how a resonant, coherent microwave signal AT-splits and shifts Rydberg-EIT lines in the absence of any noise. The AT splitting in such spectra can be used to perform an atom-based measurement of the electric field of the coherent microwave signal. Fig.~\ref{noiseEIT1}-\ref{noiseEIT13} demonstrate that BWLGN added to the system can have a profound effect on the Rydberg-EIT-AT spectra. It is seen that the details of the spectral intensity distribution of the BWLGN give rise to a wide range of cases, as to how exactly the Rydberg-EIT-AT spectra change from the noise-free case. Figs.~\ref{Efields} and \ref{Error_SNR} then show, quantitatively, the BWLGN-induced errors in the atom-based measurement of the electric field of the coherent microwave signal. These errors also strongly depend on the spectral intensity distribution of the BWLGN.


\section{Modeling the Effects of Noise}

In order to understand how the various noise sources alter the measured EIT signals, we model these effects as follows.  The effect of broadband noise on a Rydberg-atom system consists of two main contributions: (1) on-resonance transitions caused by the noise, and (2) AC Stark shift caused by the noise. The Rydberg atoms in levels $\vert 3 \rangle$ and $\vert 4 \rangle$ in Fig.~\ref{4level2}(b), which are populated by the coherent sources (lasers, coherent RF source), can transition into other Rydberg levels due to the frequency components of the noise spectrum that are resonant with transitions between Rydberg states. This process is akin to decays driven by blackbody radiation~\cite{Friedrich1990, Gallagher1994}. The usual treatment (in which the radiation field is quantized and the transition rate is obtained from Fermi's golden rule and summing over the possible field polarizations and accessible final angular-momentum states) needs to be modified so that it applies to a noise field that has a well-defined polarization and propagation direction (given by the microwave horn's geometry). Also, the black-body energy density of the field must be replaced by the setup-specific noise characteristics. We assume that, at the location of the atoms, the noise has a spectral intensity (noise intensity per frequency interval, measured in W/(m$^2\,$Hz)) of
\[I_\nu = \frac{dI}{d\nu} (\nu) \,\,\, . \]
To conform with our typical measurement scenario (i.e., the noise is applied to the atoms via a RF horn antenna and the atoms are located in the far field of the horn), we quantize the field in one dimension only (the propagation direction of the noise field) and assume a fixed noise field polarization. For the transition rate, $R_{fi}$, from an initial state $\vert i \rangle$ into a final state $\vert f \rangle$, we find
\begin{equation}
 R_{fi} = \frac{e^2}{2 \epsilon_0 \hbar^2 c} \vert {\bf{n}} \cdot \langle f \vert \hat{\bf{r}} \vert i \rangle \vert^2 I_\nu(\vert \nu_{fi} \vert) \quad,
\label{eq:gr1}
\end{equation}
where $\epsilon_0$ is the permittivity of free-space, $e$ is the elementary charge, ${\bf{n}}$ is the field-polarization unit vector, and $\nu_{fi}$ is the transition frequency (i.e., $(E_f - E_i)/h$, where $E_f$ and $E_i$ are the energies of states $\vert f \rangle$ and $\vert i \rangle$, respectively). These rates ($R_{fi}$) are in SI units and have the unit ``per atom and per second''. For the given states of interest we calculate the rates, $R_{fi}$, for the known noise spectrum $I_\nu(\nu)$. Note $R_{if}=R_{fi}$.

In the present case, the coherent microwave signal drives the transition between Rydberg states $\vert 3 \rangle$ and $\vert 4 \rangle$ in Fig.~\ref{4level2}(b). If the noise spectrum covers the transition $\vert 3 \rangle$ and $\vert 4 \rangle$, it is included in the Master equation in the form of two noise-induced bi-directional decay terms with equal rates, $R_{34}=R_{43}$, and the corresponding decay rates for the coherences that involve levels $\vert 3 \rangle$ or $\vert 4 \rangle$, or both.

For transitions $\vert 3 \rangle$ $\rightarrow$ $\vert f \rangle$ and $\vert 4 \rangle$ $\rightarrow$ $\vert f \rangle$ different from the coherently driven $\vert 3 \rangle$ $\leftrightarrow$ $\vert 4 \rangle$ transition, the noise drives transitions at rates per atom of $R_{f3}=R_{3f}$ and $R_{f4}=R_{4f}$. Note that the noise-populated levels $\vert f \rangle$ have no coherences between each other and with any of the levels $\vert 1 \rangle$ - $\vert 4 \rangle$ in Fig.~\ref{4level2}(b), because the noise-induced drive has a random phase. Hence, all levels $\vert f \rangle$ that become populated from level $\vert 3 \rangle$, due to the noise, may be lumped into a fictive level $\vert d \rangle$ (see Fig.~\ref{4level2}(b)). Similarly, all levels $\vert f \rangle$ that become populated from level $\vert 4 \rangle$, are lumped into a fictive level $\vert e \rangle$. Due to electric-dipole selection rules, there is no overlap between the levels in $\vert d \rangle$ (which become populated by the noise from $\vert 3 \rangle$) and in $\vert e \rangle$ (which become populated by the noise from $\vert 4 \rangle$).
The net rates into the fictive levels are
\begin{eqnarray}
 R_{d3} & = & \sum_{f \ne 3,4} R_{f3}  \nonumber \\
 R_{e4} & = & \sum_{f \ne 3,4} R_{f4}  \quad ,
\label{eq:gr2}
\end{eqnarray}
where $R_{d3}=R_{3d}$ and $R_{e4}=R_{4e}$. The master equation then includes equations for the level populations of the Rydberg states $\vert 3 \rangle$ and $\vert 4 \rangle$ and the fictive levels $\vert d \rangle$ and $\vert e \rangle$
\begin{eqnarray}
 \dot{\rho}_{33} & = & (other \, terms) + R_{d3} (\rho_{dd} - \rho_{33}) \nonumber \\
 \dot{\rho}_{dd} & = & - R_{d3} (\rho_{dd} - \rho_{33})  \nonumber \\
 \dot{\rho}_{44} & = & (other \, terms) + R_{e4} (\rho_{ee} - \rho_{44})  \nonumber \\
 \dot{\rho}_{ee} & = & - R_{e4} (\rho_{ee} - \rho_{44}) \quad .
\label{eq:gr3}
\end{eqnarray}
The ``other terms'' are terms that are already explained in detail in \cite{linear}. From eqs.~(\ref{eq:gr1}) and (\ref{eq:gr2}) it is seen that only transitions from the Rydberg states $\vert 3 \rangle$ or $\vert 4 \rangle$ into other levels, with transition frequencies that fall within the noise band, cause broadening of the Rydberg-EIT-AT lines. Note the master equation includes no equations for any coherences for the fictive levels (the coherences involving the fictive levels are always identical zero). The equations for the decay of coherences that involve levels $\vert 3 \rangle$ and/or $\vert 4 \rangle$ also need to be amended so that they include all $R_{3d}$-, $R_{3e}$- and $R_{34}$-terms.

The noise also induces AC shifts that are calculated based on the same field quantization model, and using second-order perturbation theory. The shifts of levels $\vert i \rangle$ $=\vert 3 \rangle$ or $\vert 4 \rangle$ are found to be
\begin{equation}
 \Delta E_i = \sum_{f \ne i} \left[ \frac{e^2 \nu_{fi}^3 \vert {\bf{n}} \cdot \langle f \vert \hat{\bf{r}} \vert i \rangle \vert^2}{h c \epsilon_0} \int_{\nu_{min}}^{\nu_{max}} \frac{I_\nu (\nu)}{\nu^2 (\nu^2-\nu_{fi}^2)} d\nu \right]\, .
\label{eq:gr4}
\end{equation}
The integration limits $\nu_{min}$ and $\nu_{max}$ are chosen wide enough that the entire noise spectrum is covered.
Note that due to the $\nu_{fi}^3$ term the signs of the transition frequencies are important (as expected).
The AC shifts of levels $\vert 3 \rangle$ and $\vert 4 \rangle$ are added into the master equation~\cite{linear} as noise-induced detuning terms. The AC shifts of other Rydberg levels, included in the model in the form of the fictive levels $\vert d \rangle$ and $\vert e \rangle$, are not important.

Comparing eqs.~(\ref{eq:gr1}-\ref{eq:gr4}) it is seen that the AC shifts are harder to calculate than the decays. For the decays, only transitions with frequencies that lie within the noise band have effects and the noise spectral density is only required at these frequencies. Typically only a few - sometimes even no - Rydberg-Rydberg transitions involving levels $\vert 3 \rangle$ or $\vert 4 \rangle$ are within the noise band. In contrast, all allowed transitions involving levels $\vert 3 \rangle$ or $\vert 4 \rangle$, including transitions with frequencies outside the noise band, are relevant in eq.~(\ref{eq:gr4}). Also, for each transition an integral over the entire noise band needs to be evaluated. For transitions within the noise band some care needs to be taken because of the pole in eq.~(\ref{eq:gr4}).

For AC-shifting transitions with frequencies $\nu_{fi}$ outside the noise band, the directions of the AC shifts of the Rydberg levels $\vert 3 \rangle$  or $\vert 4 \rangle$ due to these transitions depend on the line strengths, the signs, and the $\nu_{fi}$-values of the perturbing transitions in relation with
the noise band (see eq.~(\ref{eq:gr4})). The perturbing levels are $P$-Rydberg states for level $\vert 3 \rangle$, and $S$- and $D-$ Rydberg states for level $\vert 4 \rangle$, respectively; the frequencies of the perturbing transitions, $\nu_{fi}$, and their signs depend on the relevant quantum numbers and the quantum defects of the atom. For transitions that actually fall within the noise band, the integration range in eq.~(\ref{eq:gr4}) includes the pole; in such cases the AC-shifts depend on the details of the integrand, including any details of the noise spectrum itself. The net AC shifts of the levels $\vert 3 \rangle$  and $\vert 4 \rangle$ then result from the sum over all AC shifts, summed over all perturbing levels $\vert f \rangle$, as seen in eq.~(\ref{eq:gr4}).  The net AC shifts therefore depend on the noise spectrum and the detailed energy level structure of the atomic species used.

To evaluate Eqs.~(\ref{eq:gr1}) and (\ref{eq:gr4}), one requires the noise spectral intensity function, $I_\nu (\nu)$.  Using the free-space propagation equation given in eq.~(\ref{friis}), $I_\nu (\nu)$ is expressed as
\begin{equation}
 I_\nu (\nu) = \frac{A_{sw}^2}{x^2}\frac{c \,\,\mu_0}{2\pi} G_L(\nu) \frac{dP}{d\nu} \quad,
\label{eq:gr5}
\end{equation}
where  $x=0.342~m$ (the distance from the horn antenna to the lasers) and $G_L(\nu)$ is the linear gain for the horn antenna. Using the manufacturer's specification sheet, $G_L(\nu)$ is expressed as,
\begin{equation}
 G_L(\nu) = 10^{(15+3(\nu[{\rm{GHz}}] - 18)/8.5)/10} \quad .
\label{eq:gr6}
\end{equation}
Note the frequency ($\nu$) is entered in GHz into this equation.
The parameter $A_{sw}$ is the correction factor that accounts for the standing-wave enhancement of the field inside the cell, and as discussed above, this is determined experimentally to be 1.73.
The noise power spectral density ($\frac{dP}{d\nu}$) is given in Fig.~\ref{noise} and is normalized such that it integrates to the power (in Watts) measured (using a power meter, values are given above) at the input to the horn antenna.

The results for $I_\nu (\nu)$ for Filters 1, 2, and 3 are shown in Fig.~\ref{intenf}. These results are used in eqs.~(\ref{eq:gr1}) and (\ref{eq:gr4}) to obtain the noise-induced decay rates and AC level shifts. These are then used in eqs.~(\ref{eq:gr2}) and (\ref{eq:gr3}) and combined with standard equations for the four-level master equation~\cite{linear} to yield the coherence $\rho_{12}$ as a function of coupler-laser frequency.  The AC Stark shift caused by the coherent RF source is also included in the calculations using the Floquet analysis given in Ref. \cite{dave1}.
The four-level master equation is also amended with the coherence-free fictive ``levels'' $\vert d \rangle$ and $\vert e \rangle$ that hold the net populations driven by the noise, out of the coherently coupled levels $\vert 3 \rangle$  and $\vert 4 \rangle$.

The model EIT spectrum is then obtained by computing the Beer's absorption coefficient in the medium as a function of coupler-laser detuning, $\alpha(\Delta_C)$, for the given cell temperature (see \cite{linear} for detail). This involves an integral over the Maxwell velocity distribution in the cell~\cite{linear}, because each velocity class has its own Doppler shifts of coupler and probe beams. The ratio of input and output probe powers is then given by $\exp(-\alpha L)$, where the cell length $L=75$~mm. It is noted that, after using all experimentally available input and the computed matrix elements for all noise-driven transitions, $\langle f \vert \hat{\bf{r}} \vert i \rangle$, there is no fit parameter left to adjust the model results. Hence, absolute, fit-free agreement should be expected when comparing measured and modeled spectra of the Rydberg-EIT-AT experiments under the influence of BLWGN.

\begin{figure}[htbp]
\begin{center}
\includegraphics[scale=0.25]{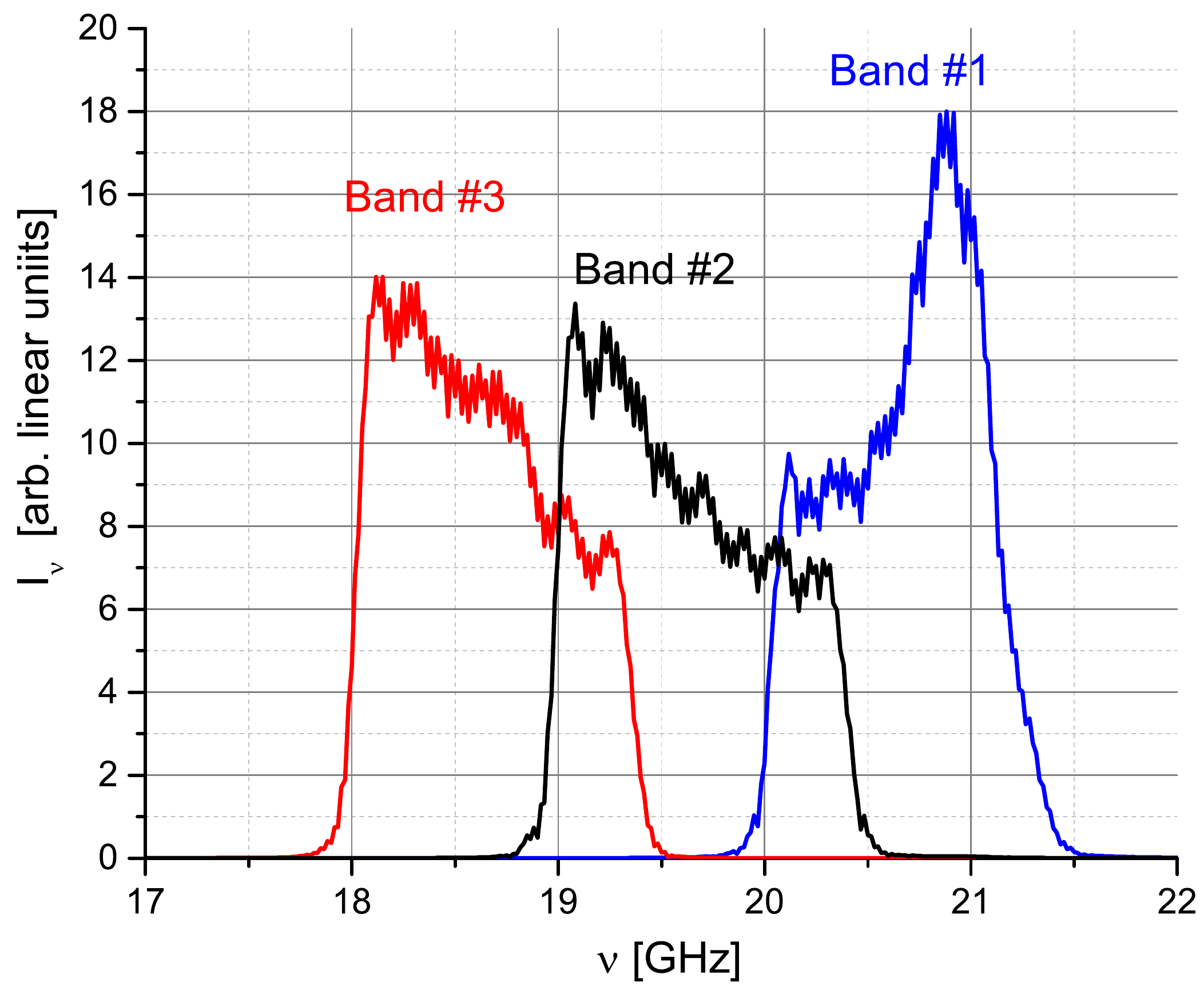}
\end{center}
\vspace{-0.8cm}
\caption{Noise spectra  $I_\nu (\nu)$ derived from measurements, the geometry of the setup, and the horn manufacturer's data on the horn's gain dependence on frequency.}
\label{intenf}
\end{figure}

In Figs.~\ref{noiseEIT1}-\ref{noiseEIT13} we show results obtained from this model (the model results are shown in plots (d)-(f) in these figures).  As a comparison, in these plots we also show peak positions from the experimental data. The squares in the figures correspond to the peaks of the experimental EIT data shown on the left-hand-side of these figures. When compared to the experimental data, we see that the model predicts the same trends in the EIT signal in the presence of noise. In that, depending on the noise source, the noise either red-shifts or blue-shifts the EIT signal.  Also, the modeled EIT signal tracks the location of the peaks as a function of $\Delta_{RF}$ very well, and when $\Omega_{RF}\rightarrow 0$, the frequency offsets are very close. In that, measured and calculated offsets are in the same directions from $\Delta_c=0$ and are usually within a couple MHz from one another. A comparison between the frequency offsets at $\Omega_{RF}= 0$ obtained from the model and those obtained from the experimental data is shown in Table~\ref{table1} (this is also indicated by the dashed lines in  Figs.~\ref{noiseEIT1}-\ref{noiseEIT3}). The largest differences are for the high noise powers (i.e., 0~dB). This is mostly due to the fact that for high noise power, the measured EIT peaks are diminished and are difficult to determine at times.  As a further comparison, we compare the experimentally determined E-field strengths to those obtained from the model based on eq.~(\ref{mage}). Fig.~\ref{compare} shows the comparison for the three filters for the high noise powers. The results in this figure show good agreement between the experimental data and the model.

The model allows for further understanding on how BLWGN influences the AT-splitting. Fig.~\ref{eitmodel} shows the AT peaks in the EIT spectra obtained from the model. Shown here are the high-power noise (0~dB attenuation) cases for the three filters. It is interesting to note the AT peaks shift in filter-specific ways relative to the no-noise cases. For large values  of $\Omega_{RF}$, Filter 3 shows only slight shifts in the peaks relative to the no-noise case. This results in the small percentage error shown in Figs.~\ref{Efields} and \ref{compare} for Filter 3.  For large values  of $\Omega_{RF}$, the Filter 2 causes both AT peaks to shift downward.  While the results for Filter 3 show that one of the peaks shifts downward and one peak shifts upward relative to the no-noise case. The shift in opposite directions for Filter 1 is what causes large errors in the E-field measurements shown in Figs.~\ref{Efields} and \ref{compare} for Filter 1.

\begingroup
\begin{table}
\caption{Comparison of frequency offset of EIT signal at $\Omega_{RF}= 0$ for various noise filters and power levels.}
\tiny
\label{table1}
\begin{center}
\begin{tabular}{|c||c|c|}\hline
& offset (MHz) & offset (MHz) \\
& Experiments & Model \\
  \hline
Filter 1 &   &   \\
-12 dB & +8  & +5   \\
-6 dB & +21  & +15  \\
0 dB & +56  & +62  \\ \hline

Filter 2 &   &   \\
-12 dB & -3  & -3   \\
-6 dB & -8  & -9   \\
0 dB & -30  & -38  \\ \hline

Filter 3 &   &   \\
-12 dB & -1  & -2  \\
-6 dB & -4  & -3  \\
0 dB & -13  & -16   \\ \hline

Filter 1/3 &    &  \\
-12 dB & +5  & +2   \\
-6 dB & +13  & +6  \\
0 dB & +35  & +23   \\ \hline
\end{tabular}
\end{center}
\end{table}
\endgroup

\begin{figure}
\centering
\scalebox{.32}{\includegraphics{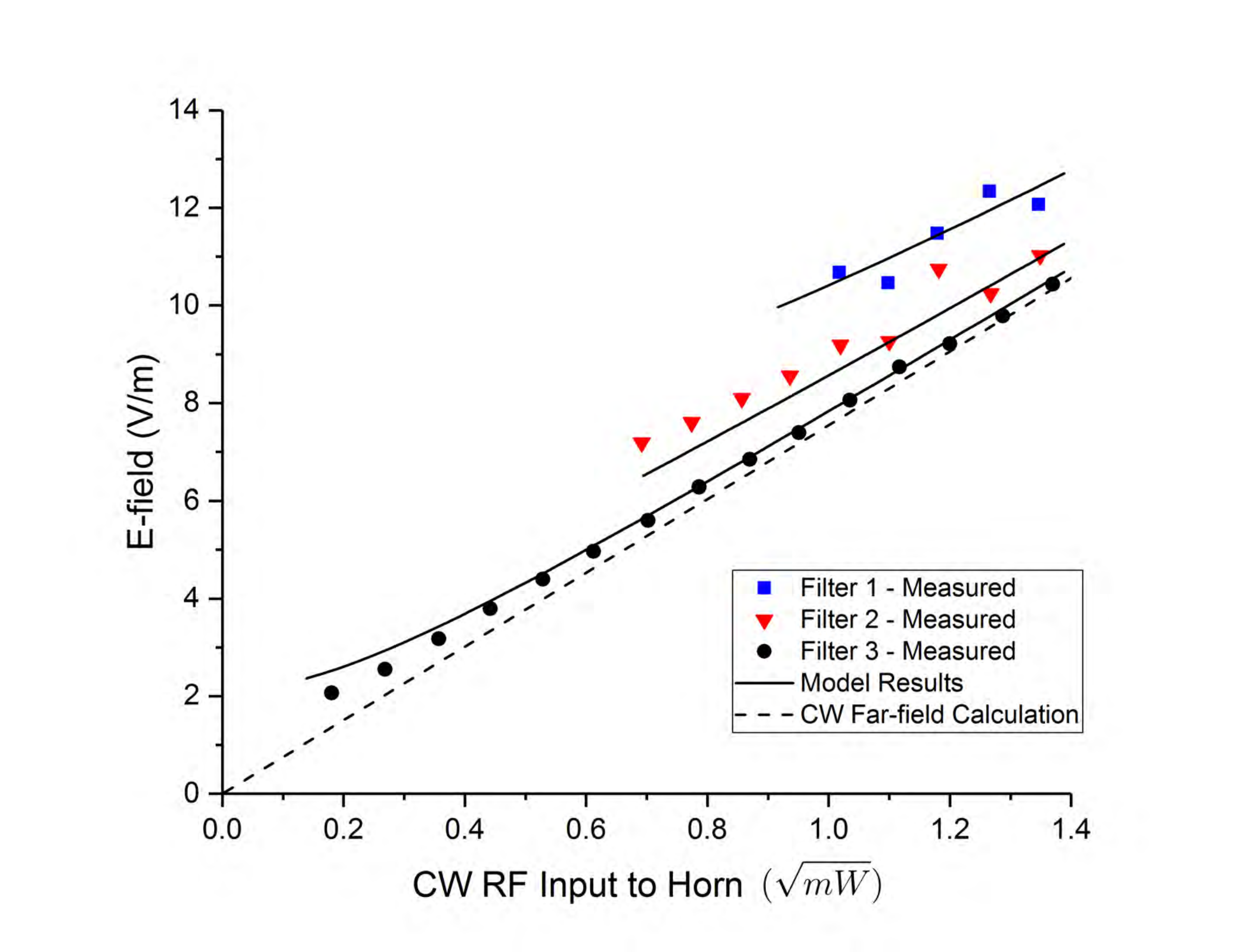}}\\
\caption{Comparison of E-field strength obtained from experimental data with those obtained from the model. The $x$-axis is the input CW RF power to the antenna.}
\label{compare}
\end{figure}

\begin{figure}
\centering
\scalebox{.32}{\includegraphics{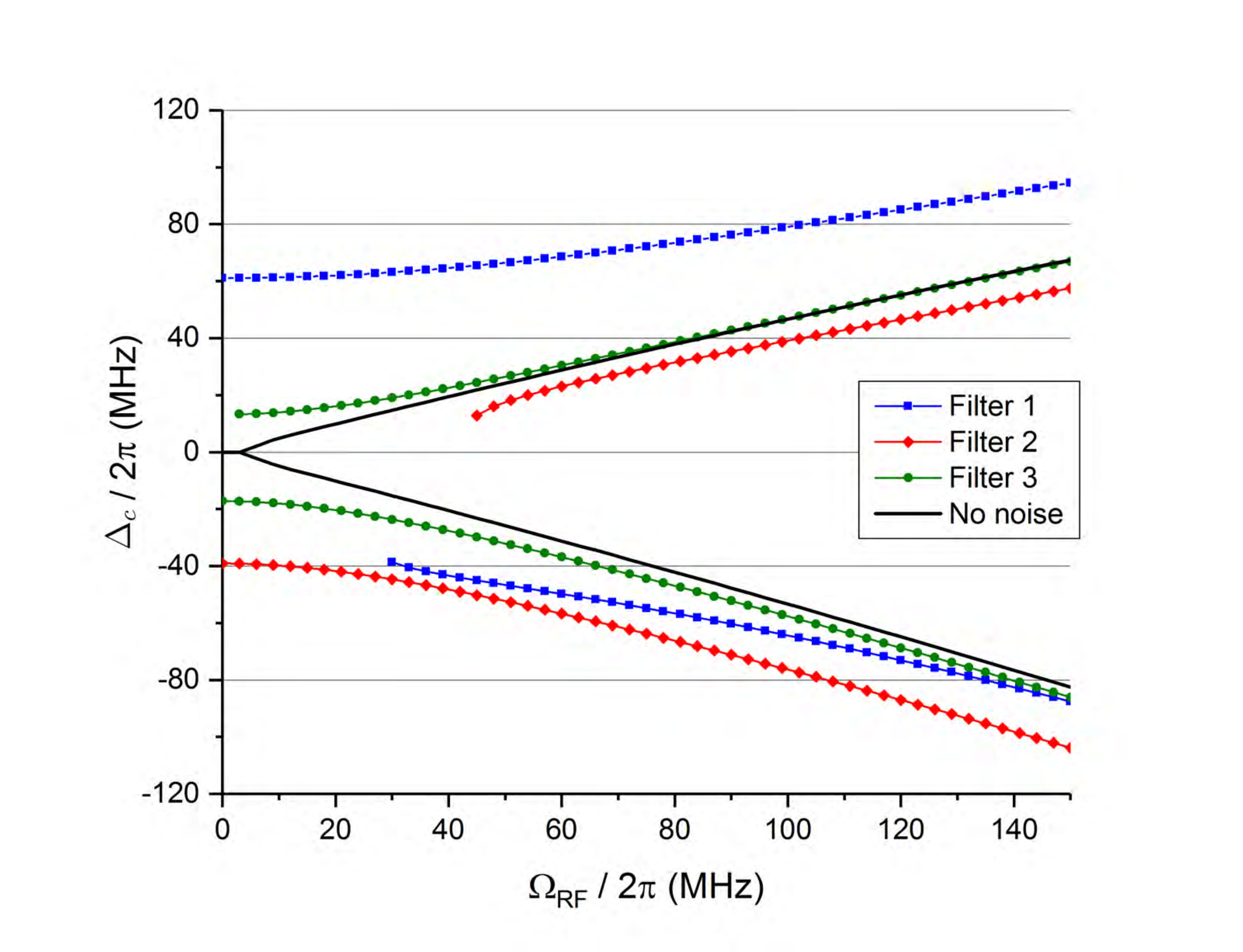}}\\
\caption{Peaks of EIT spectra obtained from the model as a function of applied coherent-RF Rabi frequency ($\Omega_{RF}$). Shown here are the high-power noise (0~dB attenuation) cases for the three filters.}
\label{eitmodel}
\end{figure}

\section{Conclusions}
The effects of band-limited white Gaussian noise on  EIT and AT splitting, when performing atom-based RF E-field strength measurements using Rydberg atomic vapor have been investigated. BLWGN has the effect of shifting (either red-shifted or blue-shifted) the peaks of the EIT lines depending on the noise conditions (band-width, center-frequency, and noise power). The BLWGN also has the effect of broadening the EIT lines. We present a model to predict these effects.  The model incorporates two contributions; (1) the on resonance transitions between Rydberg states caused by the noise spectrum (analogous to decays driven by blackbody radiation) and (2) the AC shifts caused by the noise source. The results of this model compare very well to the experimental data presented here. This indicates that these two contributions are required in a model in order to correctly predict the EIT signal in the presence of BLWGN.

The noise has the effect of modifying the EIT/AT signal from that for the coherent RF field alone, which alters the ability to measure the E-field strength in the presence of noise. The amount of deviation is a function of the noise parameters (band-width, center-frequency, and noise power).  We show relative differences of measured E-field strengths for different CSNR, where we show the the relative difference increases for decreasing CSNR.  The shifts and broadening of the EIT/AT signal caused by noise are dependent on the Rydberg states chosen for the experiment. With that said, for the Rydberg states used here, we can summarize that when the noise is red-shifted from the RF transition, it has a minimal effect on the ability to to measures CW RF \efs{}. While the blue-shifted noise (i.e., Filter 1) has the strongest effect on the \efms{}.

Furthermore, besides understanding the effects of BLWGN on E-field strength measurements, there is a need to be able to detect various noise sources in general.  In order to investigate if the EIT approach can be useful for detecting noise, we performed additional EIT experiments with various noise levels and at various temperatures (77~K to 400~K). This topic is addressed in \cite{david}.

\end{document}